\documentclass[journal,final,twoside]{IEEEtran}
\usepackage[pdftex]{graphicx}
\usepackage[cmex10]{amsmath}
\usepackage{epstopdf}
\usepackage{xcolor}
\usepackage{cite} 
\usepackage{flushend}
\usepackage{amssymb}
\usepackage{amsfonts}
\usepackage{multirow}
\usepackage{algorithm}
\usepackage{algorithmic}
\usepackage[flushleft]{threeparttable}
\usepackage[colorlinks=true,
            linkcolor=blue,
            urlcolor=blue,
            citecolor=blue,
            bookmarks=false]{hyperref}

\usepackage{subfigure}
\usepackage{amsmath}

\newtheorem{rem}{Remark}

\allowdisplaybreaks

\newcommand\articletitle{Thermal Modelling of Battery Cells for Optimal Tab and Surface Cooling Control}

\begin{document}
\title{\articletitle}
\author{Godwin~K~Peprah,~\IEEEmembership{Student Member,~IEEE}, Yicun~Huang, Torsten~Wik,~\IEEEmembership{Member,~IEEE}, Faisal Altaf, and Changfu~Zou,~\IEEEmembership{Senior Member,~IEEE}    
\thanks{This work was funded by the Swedish Electromobility Centre and the Swedish Energy Agency under grant No.~13011, the Swedish Research Council under grant No.~2019-04873, and the European Union's Horizon Europe program under the Marie Skłodowska-Curie grant agreement No.~101131278.}
\thanks{Godwin K. Peprah, Yicun Huang, Torsten Wik, and Changfu Zou are with the Department of Electrical Engineering, Chalmers University of Technology, 41296 Gothenburg, Sweden (e-mail: godwinp@chalmers.se; yicun@chalmers.se; tw@chalmers.se;  changfu.zou@chalmers.se).} 

\thanks{Faisal Altaf is with the Department of Electromobility, Volvo Group Trucks Technology, 40508 Gothenburg, Sweden (e-mail: faisal.altaf@volvo.com).}
}

\maketitle

\begin{abstract}
Optimal cooling that minimises thermal gradients and the average temperature is essential for enhanced battery safety and health. 
This work presents a new modelling approach for battery cells of different shapes by integrating Chebyshev spectral-Galerkin method and model component decomposition. As a result, a library of reduced-order computationally efficient battery thermal models is obtained, characterised by different numbers of states. These models are validated against a high-fidelity finite element model and are compared with a thermal equivalent circuit (TEC) model under real-world vehicle driving and battery cooling scenarios. Illustrative results demonstrate that the proposed model with four states can faithfully capture the two-dimensional thermal dynamics, while the model with only one state significantly outperforms the widely-used two-state TEC model in both accuracy and computational efficiency, reducing computation time by 28.7\%. Furthermore, our developed models allow for independent control of tab and surface cooling channels, enabling effective thermal performance optimisation. 
Additionally, the proposed model's versatility and effectiveness are demonstrated through various applications, including the evaluation of different cooling scenarios, closed-loop temperature control, and cell design optimisation. 


\end{abstract}

\bstctlcite{BSTcontrol}

\section{Introduction} \label{sec: Intro}
\IEEEPARstart{T}{he} pursuit of carbon-neutral energy sources has emerged as a compelling and sustainable alternative to traditional non-renewable energy such as fossil fuels. This shift has been driven by the growing global commitment to combat climate change through the reduction of carbon emissions, particularly in the transportation sector \cite{cevik2022climate}. Central to achieving this cleaner and more efficient mobility solution is the adoption of electric vehicles (EVs), whose performance strongly depends on the battery pack performance. Temperature affects several aspects of the battery, including safety, electrochemical processes, charge acceptance, round-trip efficiency, power and energy capability, and lifetime \cite{ma2018temperature}. This underscores the need for a sophisticated thermal management system \cite{lin2021review} capable of controlling the battery temperature to desired values regardless of operating conditions.

Subject to various thermal boundary conditions such as liquid or air convection \cite{tete2021developments}, battery cells can be thermally controlled at different surfaces, including the electrical connection tabs (terminals), cell surfaces, or both \cite{zhao2019cool}. Prior literature studies have investigated the impact of tab and surface cooling methods on battery thermal performance.  Hunt~{\it et al.} and Zhao~{\it et al.} concluded in \cite{hunt2016surface, zhao2018modeling} that surface cooling is more effective at maintaining lower average temperatures in  lithium (Li)-ion pouch cells, especially under high current rates, due to the larger cooling area. In a case study, these authors also demonstrated that tab cooling extended the lifetime of a battery pack by a factor of three, primarily due to the more uniform temperature distribution it provides. In a follow-up study, Dondelewski~{\it et al.} compared both methods in \cite{dondelewski2020role} and found that surface cooling resulted in lower average temperatures and slower degradation but reduced usable capacity. Conversely, tab cooling produced higher average temperatures and faster degradation, yet allowed for greater usable capacity due to more effective thermal distribution. Similar results were achieved for cylindrical Li-ion cells in \cite{li2021optimal, bolsinger2019effect}, where tab cooling was shown to reduce internal temperature inhomogeneities by approximately 25\%, though it resulted in a higher average cell temperature. Furthermore, \cite{ahmad2022identifying} reported that surface cooling causes a faster drop in temperature, driven by the larger lateral surface area available for heat transfer. However, tab cooling offered improved thermal uniformity, with a recorded temperature difference of 1.9°C, compared to 3.8°C under surface cooling.

It is clear that both surface and tab cooling methods offer distinct advantages; surface cooling can enable faster heat removal and lower average cell temperatures, while tab cooling enhances temperature uniformity and thermal balance. However, to the best of our knowledge, there is currently no battery control framework in the state-of-the-art literature that systematically investigates the optimal integration of these two cooling methods for advanced battery thermal management. 


Thermal equivalent circuit (TEC) models with lumped parameters have been extensively utilized for control-oriented modelling of batteries due to their ease of implementation and computational efficiency \cite{peprah2021optimal,lin2014lumped}. However, lumped parameter TECs can only predict average and surface temperatures and their applicability is limited to cells with small Biot numbers. Physics-based models \cite{srinivasan2002analysis} result in partial differential equations (PDEs) governing the underlying heat diffusion. They can predict the spatially distributed temperature field throughout the cell but are typically implemented via computationally expensive numerical methods, such as finite element (FE) and finite difference (FD) methods, rendering them impractical for real-time control purposes. Spectral methods \cite{shen2011spectral,trefethen2000spectral} are alternative numerical methods for finding solutions to PDEs. They belong to the class of weighted residual methods, and unlike FE and FD, they make use of global (rather than local) approximating functions in the discretization of the spatial domain, rendering them computationally efficient. With this benefit in mind, the spectral method based on the Galerkin approach was adopted to develop low-order 2-dimensional (2D) thermal models for cylindrical cells in \cite{richardson2016board} and for pouch cells in \cite{hu2020control}. However, these models do not allow independent and targeted cooling control of battery tabs and surfaces, preventing their use for optimally combining tab and surface cooling for battery thermal management.

To bridge the identified research gap, we propose new 2D battery thermal models that stem from the decomposition of the particular solution of the non-homogeneous PDE-based 2D heat equation. Unlike \cite{richardson2016board,hu2020control}, which used a lumped particular solution to capture all sides of the cell in one go, our model decomposes the particular solution into four components, each representing a distinct side of the battery cell. This method and its resulting model formulation enable us to obtain independent control signals for each side of the battery, including the tabs and surfaces. 

For a given battery system and usage profile, the choice of the most suitable cooling scenario can significantly impact its key characteristics. Hence, we conduct a model-based analysis to study the performance trade-offs amongst different cooling scenarios. Additionally, we demonstrate the application of the model to study thermal gradients under various C-rates, through a simple feedback temperature control strategy based on multiple proportional-integral (PI) controllers. Finally, using cylindrical cells as an example, we conduct a model-based study on the influence of varying dimensions on battery thermal performance.

The novel models can accurately predict the spatially resolved full temperature distribution throughout battery cells of any geometry (cylindrical, pouch, and prismatic) with higher computational efficiency compared to TEC models. Hence, it can be readily used for real-time optimal control in battery management systems, particularly for the management of the cooling effort applied on battery tabs and surfaces. It can be an efficient simulation tool for understanding battery thermal dynamics to guide the design of next-generation cooling systems and battery cells.

The remainder of this article is organized as follows. Section~\ref{Sec: Overview of Therm Mdl} details the problem underlying the battery thermal modelling. Section~\ref{Sec: Model reformulation} presents the proposed modelling framework for joint tab and surface cooling control. Section~\ref{Sec: simulations} presents the model validation and evaluation results. Section~\ref{Sec: model-based} showcases the efficacy of the model through various model-based applications and analyses. Finally, Section~\ref{Sec: conclusions} concludes the work.

\IEEEpubidadjcol

\section{Overview of the PDE-based thermal model} \label{Sec: Overview of Therm Mdl}
\begin{figure}[h!]
 \centering
 \includegraphics[width=1.0\columnwidth]{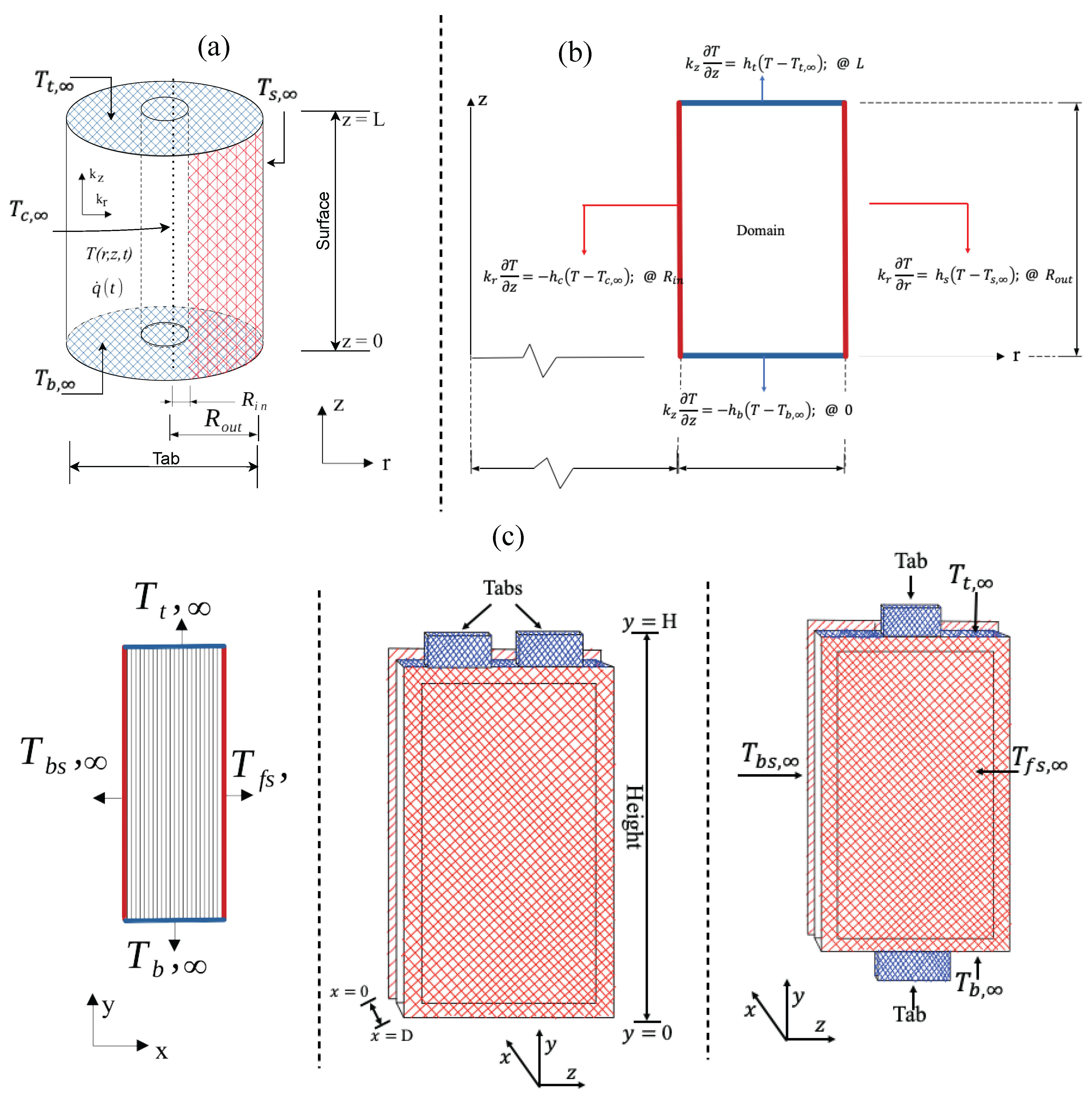}
 \caption{Schematics of different shapes of battery cells for 2D thermal modelling, with the blue and red-shaded areas representing tab and surface cooling channels, respectively. (a) A cylindrical cell cooled on the top and bottom tabs and part of the curved surface. (b) Illustration of the boundary conditions of \eqref{Eqtn: BC 1 2} for cylindrical battery cell modelling. (c)  A pouch cell whose tabs can either be on the top, bottom, or both sides.} 
 \label{fig: Battery_diag}
 \end{figure}

Battery cells of cylindrical and pouch types are widely used for electrical energy storage, as schematized in Figs.~\ref{fig: Battery_diag}(a)-(c). Considering the symmetric structure along the radial ($r$) direction of a cylindrical cell, and by assuming the same thermal conductivity in $y$- and $z$-directions of a pouch cell, the temperature distribution of both cell types can be modelled in 2D. For completeness, this section briefly reviews the first principle PDE-based models for battery thermal dynamics, including mathematical formulations as well as their underlying assumptions and considerations.


\subsection{Cylindrical cell modelling}
For the cylindrical cell, the governing equation for temperature $T(r,z,t)$ at time $t$ and position $(r,z)$ can be represented by the following 2D boundary value problem \cite{hahn2012heat}:
\begin{equation}
\begin{split}
   & \rho c_p \frac{\partial T(r,z,t)}{\partial t} - k_r \frac{\partial^2 T(r,z,t)}{\partial r^2} \\ 
   - k_r & \frac{\partial T(r,z,t)}{r \partial r}  - k_z \frac{\partial^2 T(r,z,t)}{\partial z^2} = q(t),
    \end{split}
   \label{Eqtn: GEqn}
   \end{equation}
subject to the non-homogeneous convection Robin boundary conditions given by
 \begin{subequations} \label{Eqtn: BC 1 2}
\begin{align}
  k_r\frac{\partial T}{\partial r} & = h_s\left(T-T_{s,\infty}\right) \:\:\:\text{ at } r = R_\text{out}, \\
   k_r\frac{\partial T}{\partial r} & = -h_c\left(T-T_{c,\infty}\right) \text{ at }  r = R_\text{in}, 
    \label{Eqtn: BC 1 2Db} \\
    k_z\frac{\partial T}{\partial z}  & = h_t\left(T-T_{t,\infty}\right) \:\:\: \: \text{ at } z = L, 
    \label{Eqtn: BC 1 2Dc}\\
   k_z\frac{\partial T}{\partial z} & = -h_b\left(T-T_{b,\infty}\right) \text{ at } z = 0,
   \label{Eqtn: BC 1 2D}
   \end{align}
\end{subequations}
where $\rho$, $c_p$, $k$, $h$, and $T_\infty$ represent the volume-average density, volume-average specific heat capacity, thermal conductivity, convection coefficient, and fluid free-stream temperature, respectively. $R_\text{in}$ and $R_\text{out}$ are the cylindrical cell's inner and outer radius, and $L$ is the length. The subscripts $s$, $c$, $t$, and $b$ denote the cell's surface, core, top, and bottom, respectively. 
Equation \eqref{Eqtn: BC 1 2} is illustrated in Fig.~\ref{fig: Battery_diag}(b), where the convective heat transfer is presumed to occur on the external surfaces, and the heat transfer coefficient and fluid free-stream temperatures may differ for each surface. The cooling occurring in the core area is considered as zero, leading to the corresponding convection coefficient $h_c = 0$ in \eqref{Eqtn: BC 1 2Db}. 

The cylindrical cell's multi-layer composition is considered a homogeneous solid with anisotropic thermal conductivities along $r$- and $z$-directions. Due to the spirally wound structure of the electrodes of the cylindrical cell, the thermal conductivity $k_r$ would differ slightly from the inner to the outer radius but is assumed constant in this study for simplicity. The heat generation rate $q(t)$ in \eqref{Eqtn: GEqn} is assumed to be uniformly distributed in space.

For cylindrical cells, the tabs placed on the top and bottom sides can have different sizes. Existing designs make the tab's area much smaller than the cell's top/bottom area. It has been demonstrated that these designs of cylindrical cells are not efficient for battery cooling merely from the tabs  \cite{li2021optimal,tranter2020communication, lee20103d}. In this regard, this work considers the entire area of the top/bottom as the tab, forming the so-called all-tab or tab-less cell design, as illustrated in Figs.~\ref{fig: Battery_diag}a and b.  Correspondingly, the thermal convection coefficients $h_t$ and $h_b$ in \eqref{Eqtn: BC 1 2} are affected by the entire top and bottom areas, respectively, in addition to the applied cooling media.

\subsection{Pouch cell modelling}
The thermal model for pouch cells illustrated in Fig.~\ref{fig: Battery_diag}(c) can be derived in a similar fashion as for the cylindrical cell.
The thermal conductivity in the $y$-direction is assumed to be identical to that of the $z$-direction because of the stacked structure of the pouch cell. By considering the pouch cell's multi-layer structure as a homogeneous solid with anisotropic thermal conductivities $k_x$ in the $x$-direction and $k_y$ in the $y$-direction, the governing heat equation in Cartesian coordinates is given by the 2D boundary value problem \cite{hahn2012heat}
\begin{equation} 
\label{Eqtn: GEqn_normalized}
     \rho c_p \frac{\partial T(x,y,t)}{\partial t} - k_x\frac{\partial^2 T(x,y,t)}{\partial x^2} - k_y \frac{\partial^2 T(x,y,t)}{\partial y^2} = q(t),
\end{equation}
subject to the non-homogeneous convection Robin boundary conditions expressed as
 \begin{subequations} 
   \label{Eqtn: BC pouch 2D}
\begin{align}
 k_x \frac{\partial T}{\partial x} & = h_{fs}\left (T-T_{fs,\infty}\right) \:\:\: \text{ at } x = D, \\
  k_x \frac{\partial T}{\partial x}& = -h_{bs}\left (T-T_{bs,\infty}\right) \text{ at } x = 0,
     \\
   k_y \frac{\partial T}{\partial y}  & = h_t\left (T-T_{t,\infty}\right) \:\:\:\:\:\: \:\text{ at }  y = H, \\
  k_y \frac{\partial T}{\partial y} & = -h_b\left (T-T_{b,\infty}\right) \:\:\: \text{ at } y = 0,
   \end{align}
\end{subequations}
where $D$, $H$, the subscripts $fs$, and $bs$ represent the pouch cell's width, height, front side, and back side, respectively.


\section{Control-oriented battery thermal modelling} \label{Sec: Model reformulation}
In this section, we demonstrate the control-oriented modelling process for the thermal behaviour of cylindrical cells. The model for pouch cells, which can be derived in a similar fashion, is introduced briefly afterwards.

\subsection{Model non-dimensionalization}\label{Sec:Non-dimensionalization}
To facilitate model approximation, reduction, and the subsequent analyses, we first remove all physical dimensions involved in the model's governing equation \eqref{Eqtn: GEqn}. Specifically, new coordinate variables are introduced along the radius and height directions
\begin{subequations}\label{eq:CoordinateChange}
\begin{align}
 \label{Eq: z coord change}
    \tilde{r} =\:& \alpha r - \frac{R_\text{out} + R_\text{in}}{R_\text{out}-R_\text{in}},  \\
    \tilde{z} = \:&  \beta z - 1,
\end{align}
\end{subequations}
where $\alpha = 2/({R_\text{out} - R_\text{in}})$ and $\beta = 2/L$. As a result, both $\tilde{r}$ and $\tilde{z}$ fall in the range of $[-1,1]$. 
Replacing all $r$ and $z$ by $\tilde{r}$ and $\tilde{z}$, respectively, the initial 2D thermal model for cylindrical cells is scaled and becomes
\begin{equation} 
\label{Eq: New govern eqtn 2D}
\begin{split}
 & \rho c_p \frac{\partial T(\tilde{r},\tilde{z},t)}{\partial t} - \alpha^2 k_r \frac{\partial^2 T(\tilde{r},\tilde{z},t)}{\partial \tilde{r}^2} \\
- \gamma &\frac{\partial T(\tilde{r},\tilde{z},t)}{\partial \tilde{r}} - \beta^2k_z \frac{\partial^2 T(\tilde{r},\tilde{z},t)}{\partial \tilde{z}^2} = q(t),  
 \end{split}
\end{equation}
subject to the boundary conditions
\begin{subequations} 
   \label{Eq: scaled BCs}
\begin{align}
   h_sT  - \alpha k_r \frac{\partial T}{\partial \tilde{r}} &= u_s \:\:\:\:\:\:\:\:\:\: \text{at} \:\: \tilde{r}= 1, \\
  -h_cT - \alpha k_r \frac{\partial T}{\partial \tilde{r}} &= u_c \:\: \:\:\:\:\:\:\:\:  \text{at} \:\: \tilde{r}= -1, \\
  h_t T  - \beta k_z \frac{\partial T}{\partial \tilde{z}} &=  u_t \:\:\:\:\:\:\:\:\:\: \text{at} \:\: \tilde{z}= 1, \\
  -h_b T - \beta k_z \frac{\partial T}{\partial \tilde{z}} &=  u_b \:\:\:\:\:\:\:\:\:\:\: \text{at} \:\: \tilde{z}= -1,
   \end{align}
\end{subequations}
where $\gamma =  \alpha^2 k_r/(1+ \Tilde{r} + \alpha R_\text{in})$. 
The cooling power applied per unit area is denoted by $u$, and their values at the surface, core, top and bottom sides of the cell are given, respectively, by
\begin{subequations} 
 \label{Eq: new controls}
\begin{align} 
 u_s(t) =\:& h_s(t)T_{s,\infty}(t), \:\:\:\:\:\:\: u_c(t) = -h_c(t)T_{c,\infty}(t) = 0, \\ u_{t}(t)=\:& h_t(t)T_{t,\infty}(t), \:\:\:\:\:\:\:\: u_{b}(t) = -h_b(t)T_{b,\infty}(t).
\end{align}
\end{subequations} 
The above cooling power can be considered as the dynamic thermal system's input. The coolant's temperature and flow rate can influence each of these inputs.  Furthermore, \eqref{Eq: new controls} can be integrated with a flow model to better capture the convective cooling effects and dynamic interactions between the coolant and the battery system under various cooling scenarios.

\subsection{The overall solution}\label{Sec:Overall-solution}
According to the boundary lifting principle \cite{farlow1993partial,nagle1996fundamentals,trefethen2000spectral}, the overall solution to the boundary value problem formulated in \eqref{Eq: New govern eqtn 2D}--\eqref{Eq: new controls} is composed of the homogeneous solution $T_h$ and the particular solution $T_p$. 
Namely, there exists 
\begin{equation}
\label{Eq: Gen soln}
T(\Tilde{r}, \tilde{z},t) = T_h(\Tilde{r}, \tilde{z},t) + T_p(\Tilde{r}, \tilde{z},t).
\end{equation}
Here, $T_h$ satisfies the modified problem for the cylindrical cell with the governing equation
\begin{equation} 
 \rho c_p \frac{\partial T_h}{\partial t} - \alpha^2 k_r \frac{\partial^2 T_h}{\partial \tilde{r}^2} -\gamma \frac{\partial T_h}{\partial \tilde{r}} -  \beta^2k_z \frac{\partial^2 T_h}{\partial \tilde{z}^2} = 
 q^*,
   \label{Eq: modified problem}
\end{equation}
and the homogeneous boundary conditions
\begin{subequations} 
   \label{Eq: homo BCs}
\begin{align}
   h_sT_h  - \alpha k_r \frac{\partial T_h}{\partial \tilde{r}} &= 0 \:\:\:\:\:\:\:\:\:\: \text{at} \:\: \tilde{r}= 1,
   \label{Eq: homo BCs_a} \\
  -h_cT_h - \alpha k_r \frac{\partial T_h}{\partial \tilde{r}} &= 0 \:\:\:\:\:\:\:\:\:\:  \text{at} \:\: \tilde{r}= -1, 
  \label{Eq: homo BCs_b} \\
  h_t T_h  - \beta k_z \frac{\partial T_h}{\partial \tilde{z}} &=  0 \:\: \:\:\:\:\:\:\:\ \text{at} \:\: \tilde{z}= 1, 
  \label{Eq: homo BCs_c}\\
  -h_b T_h - \beta k_z \frac{\partial T_h}{\partial \tilde{z}} &=  0 \:\:\:\:\:\:\:\:\:\:\: \text{at} \:\: \tilde{z}= -1,
  \label{Eq: homo BCs_d}
   \end{align}
\end{subequations}
where $q^*$ in \eqref{Eq: modified problem} is defined as
\begin{align}
    q^* \triangleq  q - \left( - \alpha^2 k_r \frac{\partial^2 T_p}{\partial \tilde{r}^2} -\gamma\frac{\partial T_p}{\partial \tilde{r}} -  \beta^2k_z \frac{\partial^2 T_p}{\partial \tilde{z}^2} \right). 
    \label{Eq:q_homo}
\end{align}
The particular solution for the locally distributed temperature $T_p$ in \eqref{Eq: Gen soln} satisfies the original boundary conditions defined in \eqref{Eq: scaled BCs}, leading to 
\begin{subequations} 
   \label{Eq:ScaledBCs_Tp}
\begin{align}
   h_sT_p  - \alpha k_r \frac{\partial T_p}{\partial \tilde{r}} &= u_s \:\:\:\:\:\:\:\:\:\: \text{at} \:\: \tilde{r}= 1, 
   \label{surf modified particular} \\
  -h_cT_p - \alpha k_r \frac{\partial T_p}{\partial \tilde{r}} &= u_c \:\:\:\:\:\:\:\:\:\:  \text{at} \:\: \tilde{r}= -1, \label{core modified particular}\\
  h_t T_p  - \beta k_z \frac{\partial T_p}{\partial \tilde{z}} &=  u_t \:\: \:\:\:\:\:\:\:\ \text{at} \:\: \tilde{z}= 1,
  \label{top modified particular}\\
  -h_b T_p - \beta k_z \frac{\partial T_p}{\partial \tilde{z}} &=  u_b \:\:\:\:\:\:\:\:\:\: \text{at} \:\: \tilde{z}= -1.
  \label{btm modified particular}
   \end{align}
\end{subequations}
After the particular solution $T_p$ is solved from \eqref{Eq:ScaledBCs_Tp}, it can be substituted into \eqref{Eq:q_homo} to facilitate the determination of the homogeneous solution $T_h$ governed by \eqref{Eq: modified problem}--\eqref{Eq:q_homo}.

\subsection{Model reduction for the homogeneous solution}\label{Sec:T_h}
The homogeneous problem given by \eqref{Eq: modified problem}--\eqref{Eq:q_homo} can be solved using the Chebyshev spectral-Galerkin (CSG) method. Although Jacobi polynomials, such as Legendre or Chebyshev polynomials, could be effective candidates, Chebyshev functions are employed in this work because they can be readily defined to satisfy a wide range of boundary conditions \cite{shen2011spectral}, including the Robin boundary conditions encountered in this homogeneous problem. 

Within the CSG method, the homogeneous solution $T_h$ to the 2D thermal model of cylindrical battery cells 
can be approximated by a finite sum of basis functions in the form of
\begin{equation}
\begin{split}
\hat{T}_{h}(\Tilde{r}, \tilde{z},t) =   \sum^M_{m = 1} \sum^N_{n = 1} c_{mn}(t) \phi_m^{\tilde{r}}(\tilde{r}) \phi_n^{\Tilde{z}}(\tilde{z}),
\end{split}
\label{Eq: T_h approx}
\end{equation}
where $c_{mn}(t)$ are unknown expansion (solution) coefficients. $\phi_m^{\Tilde{r}}(\Tilde{r})$ is the $m$-th Chebyshev basis function along the dimension $\tilde{r}$, satisfying the boundary conditions \eqref{Eq: homo BCs_a}--\eqref{Eq: homo BCs_b}. $M$ is the number of Chebyshev basis functions, and $m\in\{1,\cdots,M\}$. Along the dimension $\tilde{z}$, $\phi_n^{\Tilde{z}}(\Tilde{z})$ is the $n$-th Chebyshev basis function satisfying the boundary conditions \eqref{Eq: homo BCs_c}--\eqref{Eq: homo BCs_d}, where $n\in\{1,\cdots,N\}$. $M$ and $N$ are tuning parameters depending on the required accuracy. 
 
Let $P_m  (\Tilde{r}) = \cos \: (m \theta^{\Tilde{r}})$ and $P_n (\Tilde{z}) = \cos \: (n \theta^{\Tilde{z}})$, with  $\theta^{\Tilde{r}} = \arccos \: (\Tilde{r})$ and $\theta^{\Tilde{z}} = \arccos \: (\Tilde{z})$. Here, $P_m(\Tilde{r})$ and $P_n(\Tilde{z})$ represent the Chebyshev polynomials of the first kind, of degrees $m$ and $n$, respectively. 
In spectral methods, to enable efficient solution computations and satisfaction of boundary conditions, neighbouring orthogonal polynomials should be used to form the basis functions. Therefore, we seek the basis functions as a compact combination of Chebyshev polynomials in the form \cite{shen2011spectral}
 \begin{subequations} 
   \label{Eq: basis def}
\begin{align}
   \phi_m^{\Tilde{r}}(\Tilde{r}) &= P_m(\Tilde{r}) + a_m^{\Tilde{r}}(\Tilde{r}) P_{m+1}(\Tilde{r}) + b_m^{\Tilde{r}}(\Tilde{r}) P_{m+2} (\Tilde{r}),\\
 \phi_n^{\Tilde{z}}(\Tilde{z}) &= P_n(\Tilde{z}) + a_n^{\Tilde{z}}(\Tilde{z}) P_{n+1}(\Tilde{z}) + b_n^{\Tilde{z}}(\Tilde{z}) P_{n+2} (\Tilde{z}),
   \end{align}
\end{subequations}
where the coefficients $a_m^{\Tilde{r}}$, $b_m^{\Tilde{r}}$, $a_n^{\Tilde{z}}$ and $b_n^{\Tilde{z}}$ are defined according to Lemma 4.3 of \cite{shen2011spectral}.

Substituting \eqref{Eq: T_h approx} for $T_h$ into \eqref{Eq: modified problem} yields the residual denoted by $R$
\begin{equation} 
R = \rho c_p \frac{\partial \hat{T}_{h}}{\partial t} - \alpha^2 k_r \frac{\partial^2 \hat{T}_{h}}{\partial \tilde{r}^2} -\gamma \frac{\partial \hat{T}_{h}}{\partial \tilde{r}} - \beta^2k_z \frac{\partial^2 \hat{T}_{h}}{\partial \tilde{z}^2} - q^{*} \neq 0.
 \label{Eq: resi}
\end{equation}
The principle of the CSG method is to force an integral of the resulting residual $R$ to zero as 
\begin{equation} 
\begin{split}
& \langle rR,\eta\rangle = \Bigg\langle r\Big[\rho c_p \frac{\partial \hat{T}_{h}}{\partial t} - \alpha^2 k_r \frac{\partial^2 \hat{T}_{h}}{\partial \tilde{r}^2} \\
& -\gamma \frac{\partial \hat{T}_{h}}{\partial \tilde{r}} - \beta^2k_z \frac{\partial^2 \hat{T}_{h}}{\partial \tilde{z}^2} - q^{*} \Big],\eta \Bigg\rangle  
= 0,
\end{split}
 \label{Eq: cheby test}
\end{equation}
where we use the notation $\langle f,\eta\rangle$ to represent the inner product of $f$ and a test function $\eta$ in the domain of $\Tilde{r}$ and $\Tilde{z}$, namely
$\langle f,\eta\rangle =  \int^{1}_{-1} \int^{1}_{-1} f(\Tilde{r},\Tilde{z})\eta(\Tilde{r},\Tilde{z})d\Tilde{r}d\Tilde{z}$.
Note that $r$ is included in \eqref{Eq: cheby test} to account for cylindrical coordinates. For the CSG method, the test function $\eta$ must belong to the same set of Chebyshev basis functions, thereby giving  $\eta =  \phi_m^{\Tilde{r}} \phi_n^{\Tilde{z}}$.

Substituting $\hat{T}_{h}$ in \eqref{Eq: cheby test} with its actual form in \eqref{Eq: T_h approx}, we have 
\begin{equation} 
\begin{split}
& \Bigg\langle r\Big[\rho c_p \frac{\partial \big(\sum^M_{m = 1} \sum^N_{n = 1} c_{mn}(t) \phi_m^{\tilde{r}} \phi_n^{\Tilde{z}}\big)}{\partial t} \\
& - \alpha^2 k_r \frac{\partial^2 \big(\sum^M_{m = 1} \sum^N_{n = 1} c_{mn}(t) \phi_m^{\tilde{r}} \phi_n^{\Tilde{z}}\big)}{\partial \tilde{r}^2} \\ 
& -\gamma \frac{\partial \big(\sum^M_{m = 1} \sum^N_{n = 1} c_{mn}(t) \phi_m^{\tilde{r}} \phi_n^{\Tilde{z}}\big)}{\partial \tilde{r}} \\
& - \beta^2k_z \frac{\partial^2 \big(\sum^M_{m = 1} \sum^N_{n = 1} c_{mn}(t) \phi_m^{\tilde{r}} \phi_n^{\Tilde{z}}\big)}{\partial \tilde{z}^2} 
- q^{*} \Big],\eta \Bigg\rangle = 0.
 \label{Eq: Res expansion}
 \end{split}
\end{equation} 
Finally, rewriting \eqref{Eq: Res expansion} in matrix-vector notations, this gives
\begin{align} 
& \Bigg\langle r\Bigg[\rho c_p  \Big[\phi_1^{\tilde{r}} \phi_1^{\Tilde{z}}, \cdots, \phi_M^{\tilde{r}} \phi_N^{\Tilde{z}} \Big]  \begin{bmatrix} \frac{d}{dt}c_{11}(t) \\ \vdots \\ \frac{d}{dt}c_{MN}(t) \end{bmatrix} \nonumber \\
& - \alpha^2 k_r \Bigg[ \frac{\partial^2  \phi_1^{\tilde{r}} \phi_1^{\Tilde{z}}}{\partial \tilde{r}^2}, \cdots, \frac{\partial^2  \phi_M^{\tilde{r}} \phi_N^{\Tilde{z}}}{\partial \tilde{r}^2}\Bigg] \begin{bmatrix} c_{11}(t) \\ \vdots \\ c_{MN}(t) 
 \end{bmatrix} \nonumber \\
 & - \gamma \Bigg[\frac{\partial \phi_1^{\tilde{r}} \phi_1^{\Tilde{z}}}{\partial \tilde{r}}, \cdots,\frac{\partial \phi_M^{\tilde{r}} \phi_N^{\Tilde{z}}}{\partial \tilde{r}} \Bigg] \begin{bmatrix} c_{11}(t) \\ \vdots \\ c_{MN}(t) 
 \end{bmatrix} \nonumber \\
 & - \beta^2k_z \Bigg[ \frac{\partial^2 \phi_1^{\tilde{r}} \phi_1^{\Tilde{z}}}{\partial \tilde{z}^2}  ,\cdots, \frac{\partial^2 \phi_M^{\tilde{r}} \phi_N^{\Tilde{z}} }{\partial \tilde{z}^2} \Bigg] \begin{bmatrix} c_{11}(t) \\ \vdots \\ c_{MN}(t) 
 \end{bmatrix} - q^* \Bigg], \eta \Bigg\rangle = 0.
 \label{Eq: mat vec notation}
\end{align} 

\subsection{Model reduction for the particular solution} \label{Sec:T_p}
\begin{rem}\label{Rem:rich-flexible}
Unlike the homogeneous solution, which often exhibits smoother spatial variations governed by the battery system's intrinsic properties, the particular solution $T_p$ may require a more flexible representation to accommodate abrupt changes or spatial gradients induced by external factors such as boundary conditions. Additionally, $T_p$ must satisfy not only the PDE \eqref{Eq: New govern eqtn 2D} but also the non-homogeneous boundary conditions \eqref{Eq: scaled BCs} and any forcing terms, such as the heat generation $q$, presented in the problem. These additional constraints necessitate a richer and more flexible approach than a straightforward finite-sum expansion as employed in the homogeneous case.    
\end{rem}

Based on {\it Remark~\ref{Rem:rich-flexible}}, we develop a highly adaptable approach for solving the particular solution $T_p$. This approach will be able to effectively handle abrupt changes and spatial gradients while satisfying all constraints imposed by \eqref{Eq: New govern eqtn 2D}--\eqref{Eq: scaled BCs}.

\vspace{0.5em} 
\subsubsection{Approximation of $T_p$ from a polynomial vector space}
As the first step of model reduction for the particular solution $T_p$, we project and approximate it from a polynomial vector space denoted by $\mathcal{V}_{rz}$. Following a similar approach as in \cite{doha2012efficient,auteri2003essential}, $\mathcal{V}_{rz}$ is defined as
\begin{equation}
\begin{split}
   \mathcal{V}_{rz} =& \left[\{\phi_m^{\Tilde{r}} (\tilde{r}), 1 \leq m \leq M\} \otimes \{\tilde{z},\tilde{z}^2 \} \right] \\ \oplus &
   \left[\{\phi_n^{\Tilde{z}} (\tilde{z}),  1 \leq n \leq N\}  \otimes \{\tilde{r},\tilde{r}^2 \} \right],
    \label{Eq: lifting space}
    \end{split}
\end{equation}
where $\otimes$ and $\oplus$ are the outer product and direct sum, respectively. 
$\tilde{z}$, $\tilde{z}^2$,  $\tilde{r}$, and $\tilde{r}^2$ are the basis vectors of $\mathcal{V}_{rz}$. 
\eqref{Eq: lifting space} explicitly couples the $\tilde{r}$- and $\tilde{z}$-dimensions by combining basis functions from one dimension with polynomials in the other. This coupling is essential because the heat generation and boundary conditions can create interactions between $\tilde{r}$ and $\tilde{z}$ that a simpler product space might not capture effectively. 

To elucidate the idea of finite-sum approximation in the vector space $\mathcal{V}_{rz}$, we start with a simple case. Consider a vector space $\mathcal{V}_y$ consisting of the basis vectors $y$ and $y^2$. This space can be written as 
\begin{equation}
    \mathcal{V}_y = \text{span}\{y,y^2\}.
    \label{Eq: span}
\end{equation}
Any function, $f(y) \in \mathcal{V}_y$ can be represented as a linear combination of these basis vectors. Specifically, we can write any such $f(y)$ using a finite-sum approximation 
%
%
\begin{equation}
    \label{Eq: example expansion}
    f(y) \approx \sum^{2}_{i = 1} p_iy^i, 
\end{equation}
where $p_i$ are the unknown coefficients, 
%
%
which can be determined using various methods, such as the least squares or spectral methods.
  
Following the idea of finite-sum approximation introduced in \eqref{Eq: span}-\eqref{Eq: example expansion}, approximating $T_p$ in the vector space $\mathcal{V}_{rz}$, i.e., in two space dimensions, leads to the expansion 
\begin{align}
 \hat{T}_p(\tilde{r}, \tilde{z},t)  & = \sum^N_{n = 1} \Big (D_{1,n}(\Tilde{z},t) \tilde{r} + D_{2,n}(\Tilde{z},t) \tilde{r}^2\Big)\phi_n^{\Tilde{z}}(\tilde{z}) \nonumber \\
 & + \sum^M_{m = 1} \Big (D_{3,m}(\Tilde{r},t) \tilde{z} + D_{4,m}(\Tilde{r},t) \tilde{z}^2\Big)\phi_m^{\Tilde{r}}(\tilde{r}), 
    \label{Eq: Tp_hat}
\end{align}
where $\hat{T}_p$ is the approximate solution of $T_p$, and $D_{1,n}, D_{2,n}, D_{3,m}$, and $D_{4,m}$ are unknown expansion coefficients. The first and second terms on the right-hand side of \eqref{Eq: Tp_hat} represent the lumped particular solutions in the vertical and horizontal directions of the battery cell, respectively. 
\vspace{0.3em} 
\begin{rem}\label{rem:model-error}
    The deviation between \(\hat{T}_p\) and \(T_p\) can arise from two primary sources: the chosen vector space along \(\tilde{z}\) and \(\tilde{r}\), and the finite-sum approximation, which depends on the tuning parameters \(N\) and \(M\). Generally, using higher-order polynomials and including more terms in the finite sum results in smaller approximation errors but increases the number of expansion coefficients that need to be determined. The trade-offs involved, as well as the applicability of this approximation method, will be quantitatively investigated in Section~\ref{Sec: simulations}.
\end{rem}


\vspace{0.5em} 
\subsubsection{Determination of the expansion coefficients}
Taking advantage of the orthogonality and approximation properties of the Chebyshev polynomials in \eqref{Eq: basis def}, we employ the CSG method to solve for the boundary conditions formulated in \eqref{Eq:ScaledBCs_Tp}. The process is similar to that for the homogenous problem, with the details given in the following. Substituting \eqref{Eq: Tp_hat} into the boundary condition at the surface side \eqref{surf modified particular}, yields the residual $R_s$ given by
 \begin{equation} 
R_s = h_s \hat{T}_p  - \alpha k_r\frac{\partial \hat{T}_p}{\partial \tilde{r}} - u_s \neq 0, \:\: \mbox{at} \:\: \tilde{r}= 1.
   \label{Eq: Tp surf 2}
\end{equation}
As the CSG method demands, we force the integral of the resulting residual $R_s$ to zero as  
 \begin{equation} 
  \label{Eq: R_s_D}
\langle R_s, \phi_i^{\Tilde{z}} \rangle_{\tilde{z}} = \Bigg\langle \bigg[h_s \hat{T}_p  - \alpha k_r\frac{\partial \hat{T}_p}{\partial \tilde{r}} - u_s \bigg], \phi_i^{\Tilde{z}}  \Bigg \rangle_{\tilde{z}} = 0, \:\: \text{at} \:\: \tilde{r}= 1, 
\end{equation}
where the notation $\langle f, \phi_i^{\Tilde{z}}\rangle_{\Tilde{z}} =  \int^{1}_{-1} f(\Tilde{z}) \phi_i^{\Tilde{z}}(\Tilde{z})d\Tilde{z}$ represents the one-dimensional inner product in the $\Tilde{z}$-direction, and $\phi_i^{\Tilde{z}}$ with $i \in\{1,\cdots,N\}$, are the test functions. Substituting $\hat{T}_p$ in \eqref{Eq: R_s_D} with its actual form in \eqref{Eq: Tp_hat}, and with $\tilde{r}= 1$, simplifies to
\begin{equation} 
\begin{split}
\sum^N_{n = 1} &\Big[ D_{1,n} (h_s - \alpha k_r) + D_{2,n} (h_s - 2 \alpha k_r)  \Big]  \Big\langle \phi_n^{\Tilde{z}}, \phi_i^{\Tilde{z}}  \Big\rangle_{\Tilde{z}} \\
 &= \Big\langle u_{s}, \phi_i^{\Tilde{z}}  \Big\rangle_{\Tilde{z}}.
 \end{split}
   \label{Eq: D1-D2 surf}
\end{equation}
Then, we adopt a similar process for the other vertical side (i.e., the core) and the horizontal sides (i.e., the top and bottom sides) of the battery cell with the boundary conditions formulated in \eqref{core modified particular}--\eqref{btm modified particular}. Consequently, we get
\begin{subequations} 
   \label{Eq: D1-D4 core-btm}
\begin{align}
\sum^N_{n = 1} &\Big[ D_{1,n} (h_c - \alpha k_r) + D_{2,n} (-h_c+ 2\alpha k_r)  \Big] \Big\langle \phi_n^{\Tilde{z}}, \phi_i^{\Tilde{z}}\Big\rangle_{\Tilde{z}} \notag \\
&=  \Big\langle  u_{c}, \phi_i^{\Tilde{z}} \Big\rangle_{\Tilde{z}}, \label{Eq: Core D} \\
   \sum^M_{m = 1}  &\Big[D_{3,m} (h_t -\beta k_z) + D_{4,m} (h_t - 2\beta k_z)  \Big]  \Big\langle \phi_m^{\Tilde{r}}, \phi_j^{\Tilde{r}} \Big\rangle_{\Tilde{r}} \notag \\
&= \Big\langle  u_{t}, \phi_j^{\Tilde{r}} \Big\rangle_{\Tilde{r}}, \label{Eq: top D}\\
  \sum^M_{m = 1}  & \Big[ D_{3,m} (h_b -\beta k_z) + D_{4,m} (-h_b + 2\beta k_z)   \Big]  \Big\langle \phi_m^{\Tilde{r}}, \phi_j^{\Tilde{r}}   \Big\rangle_{\Tilde{r}} \notag \\
&=  \Big\langle u_{b}, \phi_j^{\Tilde{r}}   \Big\rangle_{\Tilde{r}},
  \label{Eq: btm D}
   \end{align}
\end{subequations}
where the notation $\langle f, \phi_j^{\Tilde{r}}\rangle_{\Tilde{r}} = \int^{1}_{-1} r f(\Tilde{r}) \phi_j^{\Tilde{r}}(\Tilde{r}) d\Tilde{r}$, represents the one-dimensional inner product in the $\Tilde{r}$-direction, and $r$ is included in the inner product to account for cylindrical coordinates. $\phi_j^{\Tilde{r}}$ with $j \in\{1,\cdots,M\}$ are the test functions. 

For all values of $i$ and $j$ in \eqref{Eq: D1-D2 surf}--\eqref{Eq: D1-D4 core-btm}, there are $2N+2M$ unknown coefficients in $2N+2M$ equations. By solving these equations, we can uniquely determine $D_{1,n}$, $D_{2,n}$, $D_{3,m}$, and $D_{4,m}$ for all $n$ and $m$. To simplify the notations, along the cell's surface, core, top, and bottom sides, we define
\[
\begin{aligned}
    s_1 &= (h_s - \alpha k_r), & s_2 &= (h_s - 2\alpha k_r), \\
    t_1 &= (h_t - \beta k_z), & t_2 &= (h_t - 2\beta k_z), \\
    c_1 &= (h_c - \alpha k_r), & c_2 &= (-h_c + 2\alpha k_r), \\
    b_1 &= (h_b - \beta k_z), & b_2 &= (-h_b + 2\beta k_z).
\end{aligned}
\]
Then, \eqref{Eq: D1-D2 surf}-\eqref{Eq: D1-D4 core-btm} can be re-written as 
\begin{subequations} 
 \label{Eq: D's compact}
\begin{align}
  \Phi_{v}\left(D_1 s_1 + D_2 s_2\right) &= u_sS_v, \\
  \Phi_{v} \left(D_1 c_1 + D_2c_2\right) &= u_cS_v, \\
  \Phi_{h} \left(D_3t_1 + D_4t_2\right) &= u_tS_h, \\
  \Phi_{h} \left(D_3b_1 + D_4b_2\right) &= u_b S_h,
 \end{align}
\end{subequations} 
where the matrices for the vertical sides $\Phi_{v}(\Tilde{z}) \in \mathcal{R}^{N \times N}$ and the horizontal sides $\Phi_{h}(\Tilde{r})\in \mathcal{R}^{N \times N}$ are given by $\Phi_{v} = \langle \phi_n^{\Tilde{z}}, \phi_i^{\Tilde{z}}  \rangle_{\Tilde{z}}$ and $ \Phi_{h} = \langle \phi_m^{\Tilde{r}}, \phi_j^{\Tilde{r}}  \rangle_{\Tilde{r}}$, respectively. The source terms for the vertical sides $S_v(\Tilde{z}) \in \mathcal{R}^{N}$ and the horizontal sides $S_h(\Tilde{r}) \in \mathcal{R}^{M}$ are given by $S_v = \langle 1, \phi_i^{\Tilde{z}} \rangle_{\Tilde{z}}$ and $ S_h= \langle 1, \phi_j^{\Tilde{r}}  \rangle_{\Tilde{r}}$, respectively. $D_1$, $D_2$, $D_3$, and $D_4$ are vectors of the unknown coefficients, explicitly defined as 
\begin{align*}
D_1 =\:&  \begin{bmatrix} D_{1,1} & D_{1,2} & \cdots & D_{1,N} \end{bmatrix}^T, \\ 
D_2 =\:& \begin{bmatrix} D_{2,1} & D_{2,2} & \cdots & D_{2,N} \end{bmatrix}^T, \\
D_3 = \:& \begin{bmatrix} D_{3,1} & D_{3,2} & \cdots & D_{3,M} \end{bmatrix}^T, \\
D_4 = \:& \begin{bmatrix} D_{4,1} & D_{4,2} & \cdots & D_{4,M} \end{bmatrix}^T.
\end{align*}
By solving \eqref{Eq: D's compact}, the coefficient vectors can be determined
\begin{subequations}
\label{Eq: D before split}
\begin{align}
D_1 &= \frac{1}{s_1c_2 - s_2c_1} (c_2u_s - s_2u_c)\Phi^{-1}_{v} S_v,  \\
D_2 &= \frac{1}{s_1c_2 - s_2c_1} (-c_2u_s + s_1u_c)\Phi^{-1}_{v} S_v, \\
D_3 &= \frac{1}{t_1b_2 - t_2b_1} (b_2u_t - t_2u_b)\Phi^{-1}_{h} S_h, \\
D_4 &= \frac{1}{t_1b_2 - t_2b_1} (-b_2u_t + t_1u_b)\Phi^{-1}_{h} S_h.
\end{align}
\end{subequations}
As a result, the particular solution $\hat{T}_p$ in \eqref{Eq: Tp_hat} has been derived.

\vspace{0.5em} 
\subsubsection{Model decomposition for independent and targeted cooling control}

\begin{rem}\label{Rem:Existing-cooling}
Previous efforts in the battery community to derive \eqref{Eq: Gen soln} e.g., in \cite{richardson2016board,hu2020control}, utilised \eqref{Eq: Tp_hat} as their particular solution. However, \eqref{Eq: Tp_hat} in its present form does not allow independent and targeted cooling control of battery tabs and surfaces, preventing its use for optimally combining tab and surface cooling for enhanced battery thermal performance. 
\end{rem}
Different from existing works in the literature as noted in {\it Remark~\ref{Rem:Existing-cooling}}, we propose a new method to derive the particular solution $T_p$ in battery thermal modelling. This method and its resulting model formulation offer the flexibility of independent controls for each side of the cell, including the tabs and surfaces. It is achieved by pertinent model decomposition, in which the key idea is to decompose \eqref{Eq: Tp_hat} into four components, each representing a side of the cylindrical battery cell.

According to \eqref{Eq: D before split}, the coefficient vectors, $D_1$ and $D_2$, are functions of $u_s$ and $u_c$, while $D_3$ and $D_4$ are functions of $u_t$ and $u_b$. With this in mind, we can re-arrange \eqref{Eq: D before split} so that each coefficient vector is expressed into two terms as
\begin{equation}
\begin{split}
D_1= d^{s}_1 u_s + d^{c}_1 u_c, \: D_2 = d^{s}_2 u_s + d^{c}_2 u_c, \\ D_3 = d^{t}_1 u_t + d^{b}_1 u_b,\: D_4 = d^{t}_2u_t + d^{b}_2 u_b,
 \end{split}
    \label{Eq: D after split}
\end{equation}
where 
\begin{equation}
    \label{Eq: D after split actual form}
\begin{split}
d^{s}_1 = \frac{c_2}{s_1c_2 - s_2c_1}\Phi^{-1}_{v} S_v , \:\:\:\: d^{s}_2 = \frac{-c_2}{s_1c_2 - s_2c_1}\Phi^{-1}_{v} S_v , \\
d^{c}_1 = \frac{-s_2}{s_1c_2 - s_2c_1}\Phi^{-1}_{v} S_v , \:\:\:\: d^{c}_2 = \frac{s_1}{s_1c_2 - s_2c_1} \Phi^{-1}_{v} S_v, \\
d^{t}_1 = \frac{b_2}{t_1b_2 - t_2b_1} \Phi^{-1}_{h} S_h , \:\:\:\: d^{t}_2 = \frac{-b_2}{t_1b_2 - t_2b_1} \Phi^{-1}_{h} S_h , \\
d^{b}_1 = \frac{-t_2}{t_1b_2 - t_2b_1} \Phi^{-1}_{h} S_h  , \:\:\:\: d^{b}_2 = \frac{t_1}{t_1b_2 - t_2b_1} \Phi^{-1}_{h} S_h.
 \end{split}
\end{equation}
Substituting \eqref{Eq: D after split}--\eqref{Eq: D after split actual form} into \eqref{Eq: Tp_hat}, we have the new  expansion for the particular solution as 
\begin{equation}
\begin{split}
& \hat{T}_p(\tilde{r}, \tilde{z},t) = 
 \sum^N_{n = 1} \bigg[ \bigg( d^{s}_1(\Tilde{z})u_s(t) + d^{c}_1(\Tilde{z})u_c(t) \bigg)\tilde{r} 
 + \bigg(d^{s}_2(\Tilde{z})u_s(t) \\
 & + d^{c}_2(\Tilde{z})u_c(t)\bigg)\tilde{r}^2 \bigg] \phi_n^{\Tilde{z}}(\tilde{z}) 
  + \sum^M_{m = 1} \bigg[\bigg( d^{t}_1(\Tilde{r})u_t(t) + d^{b}_1(\Tilde{r})u_b(t) \bigg)\tilde{z} \\
  & + \bigg( d^{t}_2(\Tilde{r})u_t(t) + d^{b}_2(\Tilde{r})u_b(t) \bigg) \tilde{z}^2 \bigg]\phi_m^{\Tilde{r}}(\tilde{r}).
    \label{Eq: Tp_N 3 B}
\end{split}
\end{equation}
The above result can be re-organized into four components as 
\begin{equation}
\begin{split}
 \hat{T}_p(\tilde{r}, \tilde{z},t) = \:& \hat{T}_p^{s}(\tilde{r}, \tilde{z})u_s(t) + \hat{T}_p^{c}(\tilde{r}, \tilde{z})u_c(t) \\
 + & \: \hat{T}_p^{t}(\tilde{r}, \tilde{z})u_t(t) + \hat{T}_p^{b}(\tilde{r}, \tilde{z})u_b(t), 
    \label{Eq: Tp_N four sides}
 \end{split}
\end{equation}
where $\hat{T}_p^{s}$, $\hat{T}_p^{c}$, $\hat{T}_p^{t}$, and $\hat{T}_p^{b}$
represent the particular solution expansions projected onto the surface, core, top, and bottom, respectively. Specifically, these components are given by
\begin{subequations} 
\label{Eq: Tp_N four sides 2}
\begin{align} 
 \hat{T}_p^{s} =\:& \sum^N_{n = 1} \Big (d^{s}_1\tilde{r} + d^{s}_2 \tilde{r}^2\Big)\phi_n^{\Tilde{z}}, \:\: \hat{T}_p^{c} = \sum^N_{n = 1} \Big (d^{c}_1 \tilde{r} + d^{c}_2 \tilde{r}^2 \Big )\phi_n^{\Tilde{z}}, \\ \hat{T}_p^{t} =\:& \sum^M_{m = 1} \Big(d^{t}_1\tilde{z} + d^{t}_2\tilde{z}^2\Big )\phi_m^{\Tilde{r}}, \:\:  \hat{T}_p^{b} = \sum^M_{m = 1} \Big (d^{b}_1\tilde{z} + d^{b}_2\tilde{z}^2 \Big)\phi_m^{\Tilde{r}}.
\end{align}
\end{subequations} 
We recall from \eqref{Eq: new controls} that $u_c = 0$ for the cylindrical cell, thus $\hat{T}_p^{c}$ does not contribute to the particular solution. Due to this, we neglect  $\hat{T}_p^{c}$ in the sequel.

\begin{rem}\label{rem:superposition}
    The result obtained in \eqref{Eq: Tp_N four sides}--\eqref{Eq: Tp_N four sides 2} implies that for the particular problem in \eqref{Eq:ScaledBCs_Tp}, which leads to the particular solution $T_p$ featuring linearity in two spatial dimensions, the solution trajectory in response to the cooling power applied at each individual boundary satisfies the superposition principle. Additionally, the particular solution \({T}_p(\tilde{r}, \tilde{z}, t)\) at each position \((\tilde{r}, \tilde{z})\) can be expressed as a linear combination of 
    the cooling powers applied at different boundaries.
\end{rem}

\subsection{State-space model representation} \label{Sec: State Space}
Substituting the particular solution in \eqref{Eq: Tp_N four sides}--\eqref{Eq: Tp_N four sides 2} into \eqref{Eq:q_homo} and using the linearity of the partial differential operator, the term $q^*$, which is required for deriving the homogeneous solution $T_h$, can be obtained accordingly
\begin{equation} \label{Eq: q* after Tp_hat sub}
\begin{split}
q^* = q & + 
  \bigg[ \alpha^2 k_r \frac{\partial^2\hat{T}_p^{s}}{\partial \tilde{r}^2} + \gamma \frac{\partial \hat{T}_p^{s} }{\partial \tilde{r}} +  \beta^2k_z \frac{\partial^2 \hat{T}_p^{s}}{\partial \tilde{z}^2} \bigg]u_s \\
  & + \bigg[ \alpha^2 k_r \frac{\partial^2\hat{T}_p^{t}}{\partial \tilde{r}^2} + \gamma \frac{\partial \hat{T}_p^{t} }{\partial \tilde{r}} +  \beta^2k_z \frac{\partial^2 \hat{T}_p^{t}}{\partial \tilde{z}^2} \bigg ]u_t \\
& + \bigg[ \alpha^2 k_r \frac{\partial^2\hat{T}_p^{b}}{\partial \tilde{r}^2} + \gamma \frac{\partial \hat{T}_p^{b} }{\partial \tilde{r}} +  \beta^2k_z \frac{\partial^2 \hat{T}_p^{b}}{\partial \tilde{z}^2} \bigg]u_b.
    \end{split}
\end{equation}
Based on the results in Section~\ref{Sec:T_h}--\ref{Sec:T_p}, a control-oriented 2D thermal model of cylindrical battery cells is achieved. We define $X$ as the state variable, $u$ the input, $\mathcal{Y}$ the output, and $w$ the disturbance. According to \eqref{Eq: mat vec notation}, $X$ is given by
\begin{equation}
 \label{Eqtn: state vector 2D}
X = \left[ c_{11},\cdots, c_{1N}, c_{21},\cdots, c_{2N}, \cdots, c_{M1},\cdots, c_{MN}\right]^{T}.
\end{equation}

The input $u$ is composed of the cooling power applied at the different places, leading to
$u = \left[u_s, \: 0, \: u_t, \: u_b \right]^T$.
The disturbance $w$ represents the heat generation rate, namely $w = q$.  
The outputs of interest, $\mathcal{Y}$, include the temperatures at the mid-points of the surface, core, top, and bottom sides, namely,
\begin{equation} \label{Eq:Output}
    \mathcal{Y}(t) = 
\begin{bmatrix}
    \hat{T}(1,0,t) & \hat{T}(-1,0,t) & \hat{T}(0,1,t) & \hat{T}(0,-1,t)
\end{bmatrix}^T,
\end{equation}
where $\hat{T}$ is the approximate solution of 
$T$ in \eqref{Eq: Gen soln}. 
With these definitions, this dynamic thermal model can be presented in the state-space form
\begin{subequations}
\label{Eq:SS-model}
\begin{align}
    \mathcal{G} \dot{X}(t) =\:& \mathcal{A} X(t) + \mathcal{B} u(t) + \mathcal{F} w(t), 
    \label{Eq: state space 2D} \\
    \mathcal{Y}(t) =\:& \mathcal{C} X(t) + \mathcal{D} u(t) ,
       \label{Eq: output equation 2D} 
\end{align}
\end{subequations}
where with the additional term $\mathcal{D} =[\hat{T}_p^{s}, \: \hat{T}_p^{c}, \: \hat{T}_p^{t}, \: \hat{T}_p^{b}]^T$, the matrices $\mathcal{G}$, $\mathcal{A}$, $\mathcal{B}$, $\mathcal{C}$, and the vector $\mathcal{F}$ are derived from \eqref{Eq: Gen soln}, \eqref{Eq:q_homo}, \eqref{Eq: mat vec notation}, and \eqref{Eq: Tp_N four sides}--\eqref{Eq: Tp_N four sides 2} to yield
\begin{subequations}
\label{Eq:SS-model-definition} 
\begin{align} 
 \mathcal{G} =\:& \rho c_p\big\langle r \phi_m^{\Tilde{r}}\phi_n^{\Tilde{z}},\eta \big\rangle, 
 \\
\mathcal{A} =\:& \bigg\langle r\bigg[\alpha^2 k_r \frac{\partial^2 \phi_m^{\Tilde{r}}\phi_n^{\Tilde{z}}}{\partial \tilde{r}^2} + \gamma \frac{\partial \phi_m^{\Tilde{r}}\phi_n^{\Tilde{z}}}{\partial \tilde{r}} \notag \\
& + \beta^2 k_z \frac{\partial^2 \phi_m^{\Tilde{r}}\phi_n^{\Tilde{z}}}{\partial \tilde{z}^2}\bigg] , \eta \bigg\rangle,
 \label{Eqtn: A matrix 2D} 
 \\
\mathcal{B}(:,1) =\:& \bigg\langle r\bigg[\alpha^2 k_r \frac{\partial^2 \hat{T}_p^{s}}{\partial \tilde{r}^2} + \gamma \frac{\partial \hat{T}_p^{s}}{\partial \tilde{r}} + \beta^2 k_z \frac{\partial^2 \hat{T}_p^{s}}{\partial \tilde{z}^2}\bigg] , \eta \bigg\rangle,
  \label{Eqtn: B matrix1 2D} 
  \\
\mathcal{B}(:,2) =\:& \bigg\langle r\bigg[\alpha^2 k_r \frac{\partial^2 \hat{T}_p^{t}}{\partial \tilde{r}^2} + \gamma \frac{\partial \hat{T}_p^{t}}{\partial \tilde{r}} + \beta^2 k_z \frac{\partial^2 \hat{T}_p^{t}}{\partial \tilde{z}^2}\bigg] , \eta \bigg\rangle,
  \label{Eqtn: B matrix3 2D} 
  \\
\mathcal{B}(:,3) =\:& \bigg\langle r\bigg[\alpha^2 k_r \frac{\partial^2 \hat{T}_p^{b}}{\partial \tilde{r}^2} + \gamma \frac{\partial \hat{T}_p^{b}}{\partial \tilde{r}} + \beta^2 k_z \frac{\partial^2 \hat{T}_p^{b}}{\partial \tilde{z}^2}\bigg] , \eta \bigg\rangle,
  \label{Eqtn: B matrix4 2D}
\\
  \mathcal{C}(1,:) =\:&  \phi_m^{\Tilde{r}}(1)\phi_n^{\Tilde{z}}(0), \:\:
\mathcal{C}(2,:) = \phi_m^{\Tilde{r}}(-1)\phi_n^{\Tilde{z}}(0), \\
\mathcal{C}(3,:) =\:&  \phi_m^{\Tilde{r}}(0)\phi_n^{\Tilde{z}}(1),
\:\:
\mathcal{C}(4,:) = \phi_m^{\Tilde{r}}(0) \phi_n^{\Tilde{z}}(-1),
\\
   \mathcal{F} =\:& \langle r,\eta \rangle.
  \label{Eqtn: M and F matrix 2D}
 \end{align}
\end{subequations}
Up to this point, the initial PDE-based model for cylindrical battery cells governed by \eqref{Eqtn: GEqn}--\eqref{Eqtn: BC 1 2} has been reformulated and approximated by the above ODE-based model. The order of this model, which is also called the number of states, is $O=MN$. 

\subsection{Application of the modelling techniques to pouch cells} \label{Sec: Pris_Pouch}
The modelling techniques introduced in Section~\ref{Sec:Non-dimensionalization}--\ref{Sec: State Space} can also be applied to pouch battery cells. For brevity, only the key mathematical results are presented. 

As a result of non-dimensionalisation, the initial thermal model for pouch cells in \eqref{Eqtn: GEqn_normalized}--\eqref{Eqtn: BC pouch 2D} can be presented in the form
\begin{equation} 
 \rho c_p \frac{\partial T}{\partial t} - \lambda^2 k_x \frac{\partial^2 T}{\partial \tilde{x}^2} - \zeta^2k_y\frac{\partial^2 T}{\partial \tilde{y}^2} = q, 
 \label{Eq: scaled pouch govern eqtn 2D}
\end{equation}
with scaling factors $\lambda = 2/D $ and $\zeta = 2/H$. The new spatial coordinates are $\tilde{x} = 2x/D - 1$ and $\tilde{y} = 2y/H - 1$. The above dynamic equation is subject to the boundary conditions
\begin{subequations} 
   \label{Eq: scaled BCs pouch}
\begin{align}
   h_{fs}T  -\lambda k_x\frac{\partial T}{\partial \tilde{x}} &= u_{fs} \:\:\:\:\:\:\:\:\:\: \text{at} \:\: \tilde{x}= 1 \\
  -h_{bs}T -\lambda k_x \frac{\partial T}{\partial \tilde{x}} &= u_{bs} \:\:\:\:\:\:\:\:\:\: \text{at} \:\: \tilde{x}= -1 \\
  h_t T -\zeta k_y \frac{\partial T}{\partial \tilde{y}} &=  u_t \:\: \:\:\:\:\:\:\:\:\:\: \text{at} \:\: \tilde{y}= 1 \\
  -h_b T -\zeta k_y \frac{\partial T}{\partial \tilde{y}} &=  u_b \:\: \:\:\:\:\:\:\:\:\:\: \text{at} \:\: \tilde{y}= -1,
   \end{align}
\end{subequations}
where $u_{fs}$, $u_{bs}$, $u_t$, and $u_b$ represent the cooling power applied per unit area at the front, back, top, and bottom sides of the pouch cell, respectively, which are mathematically defined as
\begin{subequations} 
 \label{Eq: new controls pouch}
\begin{align} 
 u_{fs}(t) =\:& h_{fs}(t)T_{fs,\infty}(t), \:\: u_{bs}(t) = -h_{bs}(t)T_{bs,\infty}(t), \\ u_{t}(t) =\:&  h_t(t)T_{t,\infty}(t), \:\: u_{b}(t) = -h_b(t)T_{b,\infty}(t).
\end{align}
\end{subequations} 
The outputs of interest $\mathcal{Y}$, for pouch battery cells are located at the middle of each side, with the mathematical definition being the same as \eqref{Eq:Output}.

Following similar steps from \eqref{Eq: Gen soln}--\eqref{Eq: q* after Tp_hat sub}, the battery model in \eqref{Eq: scaled pouch govern eqtn 2D}--\eqref{Eq: scaled BCs pouch} can also be approximated by an ODE-based model in the state-space form of \eqref{Eq:SS-model} with the state variable defined in \eqref{Eqtn: state vector 2D} and the input $u = [u_{fs}, \: u_{bs}, \: u_t,  \: u_b]^T$. The resulting matrices and vectors for this state-space model are given by 
\begin{subequations} 
   \label{Eq: B matrix pouch}
\begin{align}
\mathcal{G} & = \rho c_p\big\langle\phi_m^{\Tilde{x}}\phi_n^{\Tilde{y}},\eta \big\rangle,
\\
\mathcal{A} &= \bigg\langle\lambda^2 k_x \frac{\partial^2\phi_m^{\Tilde{x}}\phi_n^{\Tilde{y}}}{\partial \tilde{x}^2} + \zeta^2 k_y \frac{\partial^2 \phi_m^{\Tilde{x}}\phi_n^{\Tilde{y}}}{\partial \tilde{y}^2}, \eta \bigg\rangle, 
\label{Eqtn: A matrix pouch} \\
\mathcal{B}(:\:,1) &= \bigg\langle\lambda^2 k_x \frac{\partial^2 \hat{T}^{fs}_p}{\partial \tilde{x}^2} + \zeta^2 k_y \frac{\partial^2 \hat{T}^{fs}_p}{\partial \tilde{y}^2}, \eta \bigg\rangle,
  \label{Eqtn: B matrix1 pouch} \\
  \mathcal{B}(:,2) &= \bigg\langle\lambda^2 k_x \frac{\partial^2 \hat{T}^{bs}_p}{\partial \tilde{x}^2} + \zeta^2 k_y \frac{\partial^2 \hat{T}^{bs}_p}{\partial \tilde{y}^2}, \eta \bigg\rangle,
  \label{Eqtn: B matrix2 pouch} \\
\mathcal{B}(:\:,3) &= \bigg\langle\lambda^2 k_x \frac{\partial^2 \hat{T}^{t}_p}{\partial \tilde{x}^2} + \zeta^2 k_y \frac{\partial^2 \hat{T}^{t}_p}{\partial \tilde{y}^2}, \eta \bigg\rangle, 
  \label{Eqtn: B matrix3 pouch} \\
  \mathcal{B}(:\:,4) &= \bigg\langle\lambda^2 k_x \frac{\partial^2 \hat{T}^{b}_p}{\partial \tilde{x}^2} + \zeta^2 k_y\frac{\partial^2 \hat{T}^{b}_p}{\partial \tilde{y}^2}, \eta \bigg\rangle,
  \label{Eqtn: B matrix4 pouch}
  \\
  \mathcal{C}(1,:) & =  \phi_m^{\Tilde{x}}(1)\phi_n^{\Tilde{y}}(0), \:\:
\mathcal{C}(2,:) = \phi_m^{\Tilde{x}}(-1)\phi_n^{\Tilde{y}}(0), \\
\mathcal{C}(3,:) & =  \phi_m^{\Tilde{x}}(0)\phi_n^{\Tilde{y}}(1),
\:\:
\mathcal{C}(4,:) = \phi_m^{\Tilde{x}}(0) \phi_n^{\Tilde{y}}(-1),
\\
 \mathcal{F} & = \langle 1,\eta \big\rangle,
 \label{Eqtn: G and F pouch}
   \end{align}
\end{subequations} 
where $\hat{T}^{fs}_p$ and $\hat{T}^{bs}_p$ represent the particular solutions of the front- and back-side expansions, respectively. 

 At higher C-rates, localized heat generation at the battery tabs can significantly influence the internal temperature distribution of the cell \cite{xie2020improved}. Future work will explore modelling approaches that treat the tabs and cell body as thermally coupled regions using a contact resistance framework. This will enable a more spatially resolved thermal analysis and better capture the impact of tab heating on overall cell performance.

Due to the internal structure of prismatic cells, significant thermal gradients can develop in all space dimensions ($x$, $y$, and $z$), requiring a 3D thermal model for better accuracy. However, the underlying solution processes for the prismatic cell is equivalent to that of the cylindrical and pouch cells already discussed.

\section{Model validation and evaluation} \label{Sec: simulations}
In this section, we evaluate the developed spatially distributed thermal models for cylindrical battery cells in \eqref{Eq:SS-model}--\eqref{Eq:SS-model-definition} and for pouch cells in \eqref{Eq:SS-model} and \eqref{Eq: B matrix pouch}. In addition to the computational efficiency, the evaluation criteria include model accuracy for the maximum temperature $T_{\max}$, mean temperature $T_\text{mean}$, maximum thermal gradient along the radius direction $\delta T^r_{\max}$, maximum thermal gradient along the axial direction $\delta T^z_{\max}$, and mean thermal gradient $\delta T^r_{\text{mean}}$. Then, we incorporate these models in a feedback controller for battery cooling and investigate the effectiveness of various tab and surface cooling scenarios. Additionally, using cylindrical cells as an example, we study the influence of varying dimensions on battery thermal performance.

\subsection{Benchmarks}\label{sec:benchmark}
To comprehensively and systematically assess the proposed modelling approach, we introduce two types of benchmarks. The first type is the original PDE models, i.e., \eqref{Eqtn: GEqn}--\eqref{Eqtn: BC pouch 2D}, solved by the finite element method (FEM). Results from these models will serve as ground truth in comparative studies. The second type of benchmark is TEC models that are widely studied in the literature  \cite{lin2012online,lin2014lumped,peprah2021optimal,zou2018power} and have been utilised in commercial battery management systems (BMSs). This kind of model features lumped parameters and two states in the form
\begin{subequations} \label{Eq: TEC model}
    \begin{align}
        C_c \dot{T}_c(t) &= q(t) + \frac{T_s(t)-T_c(t)}{R_c}, \\
        C_s \dot{T}_s(t) &= \frac{T_\infty-T_s(t)}{R_u} + \frac{T_s(t)-T_c(t)}{R_c},
    \end{align}
\end{subequations}
where $T_c$ and $T_s$ represent the core and surface temperatures, respectively. $R_c$ is the heat conduction resistance, which quantifies the heat exchange between the core and surface. $R_u$ is the convection resistance, which measures the convection cooling along the battery surface and is dependent on the pack geometry, coolant type, and flow rate. $C_c$ and $C_s$ indicate the heat capacity of the core and surface (casing), respectively.

To utilize the TEC model in \eqref{Eq: TEC model}, we first need to identify the unknown parameters, i.e., $C_c$, $C_s$, $R_c$ and $R_u$, for given battery cells and cooling scenarios. We follow the parameterisation scheme proposed in \cite{lin2014lumped}, which was designed using the least squares algorithm and the measurable current $I$ (or $q$), $T_s$, and $T_\infty$. 
According to \cite{lin2014lumped}, only three of the four parameters can be identified from these measurements, and one parameter must be assumed. In fact, the heat capacity of the aluminium casing $C_s$ can be calculated from its specific heat capacity and dimensions. 

\subsection{Specification and simulation setup} \label{Sec: Sim setup}
To demonstrate the modelling for cylindrical cells, we choose a large format lithium-iron-phosphate (LFP) $45$Ah cell and take the corresponding model parameters from \cite{roscher2015synchronisation}. Specifically, for dimension-related parameters, we have $L$ = 198~mm, $R_{\text{out}}$ = 32~mm, $R_{\text{in}}$ = 4~mm \cite{roscher2015synchronisation}, and for thermo-physical parameters, we use $\rho$ = 2118~kgm$^{-3}$, $c_p$ = 795~Jkg$^{-1}$K$^{-1}$ $k_r$ = 0.67~Wm$^{-1}$K$^{-1}$, and $k_z$ = 66.6~Wm$^{-1}$K$^{-1}$ \cite{fleckenstein2013thermal,lin2014lumped}. For the TEC model of this cylindrical cell, $C_s$ is found to be 48.35~JK$^{-1}$. Regarding $R_c$ and $R_u$, it is important to note that different sets of parameters will be obtained, depending on the employed cooling scenario. For brevity, only the identified parameters for surface cooling are presented here. After validation, we obtained the parameter values as $C_c$ = 1079.6~JK$^{-1}$, $R_c$ = 0.65~KW$^{-1}$, and $R_u$ = 0.08~KW$^{-1}$.

In the simulation-based study of the proposed battery models and their benchmarks, the initial temperatures, i.e., $T(r,z,0)$, $T(x,y,0)$, $T_c(0)$, and $T_s(0)$, and the fluid free-stream (ambient) temperature $T_\infty$ are all chosen to be 15~$^\circ$C.  To facilitate the study without losing generality, the volumetric heat generation rate $q$ is given by the heat generation model proposed by Bernardi~{\it et al.} \cite{bernardi1985general}
\begin{equation}
q(t) = I(t)\left[V(t)-V_\text{ocv}(t)\right]/V_b(t),
\label{Eq: q Bernadi}
\end{equation}
where $I, V, V_\text{ocv}$, and $V_b$ are the battery current, terminal voltage, open-circuit voltage, and volume, respectively. Here, the trajectories of $I$ and $V$ depend on the battery usage. We select the worldwide harmonised light vehicle test procedure (WLTP) \cite{mock2014wltp}, which represents a wide range of vehicle driving conditions on urban, suburban, and highway roads, and scale it so that 
the maximum current becomes 1 C-rate for heat generation. The modified WLTP (mWLTP) heat generation profile is shown in Fig.~\ref{Fig: FEM_MaxError}(a). $V_\text{ocv}$ depends on the state of charge (SoC), and the relationship between $V_\text{ocv}$ and SoC has been determined using a lookup table for a typical LFP cell in \cite{li2013comparative}. 

Depending on where the cooling action is imposed, five different cooling scenarios are investigated, including surface cooling (SC), bottom tab cooling (bTC), bottom tab and surface cooling (bTSC), bottom and top tabs cooling (btTC), and all-tabs and surface cooling (aTSC). aTSC has all sides of the battery cell exposed to cooling and is representative of immersion cooling \cite{roe2022immersion} where the cell is immersed in a dielectric fluid. 
Forced convection liquid cooling (for instance, using a cooling plate) via water or glycol is considered to apply on areas where tab or surface cooling occurs. This typically has a large convection coefficient $h$, which is set to be 400~Wm$^{-2}$K$^{-1}$, according to \cite{tete2021developments}. For areas not exposed to active cooling, mild air convection occurs, resulting in a small convection coefficient specified as $h$ = 30~Wm$^{-2}$K$^{-1}$.

All the simulations are conducted on a MacBook Pro with a 2.6GHz 6-Core Intel Core i7 processor and 16GB RAM. The first type of benchmark model is solved by the finite element method with an extremely fine mesh of 3610 triangular elements and is implemented in COMSOL Multiphysics v6.0 \cite{multiphysics1998introduction}. 
Our developed CSG models and their TEC alternatives are implemented in \textsc{MATLAB} R2023b.
For simplicity in simulating the CSG models, the numbers of basis functions in the radial and axial directions, as introduced in \eqref{Eq: basis def}, are chosen to be equal, i.e., 
$M = N$, and the model order $O= N^2$. 

\subsection{Results of validation against the FEM model} \label{Sec: Validation against FEM}
\begin{figure}[ht]
    \centering  
    \includegraphics[width=1.0\linewidth]{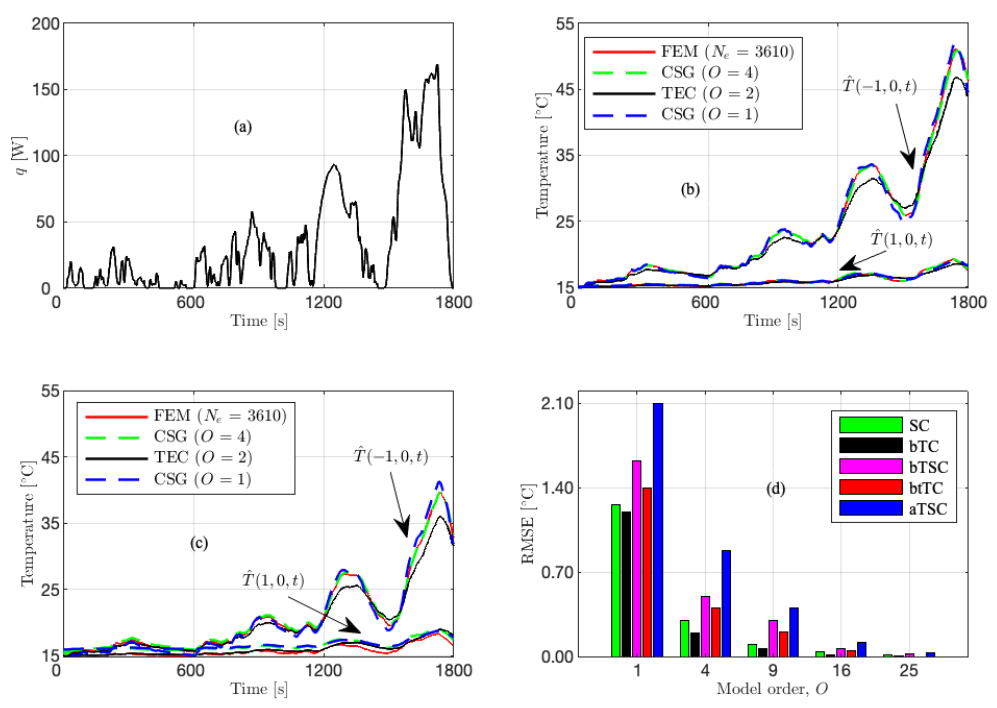}
    \caption{Validation of the developed CSG thermal model for cylindrical battery cells. (a) Heat generation profile of the mWLTP. (b) Comparison of the CSG model of orders $O$ = 1 and 4, against the FEM model with the element number $N_e$ = 3610 and TEC model, under surface cooling (SC). Comparisons are done for the surface and core mid-point temperatures, i.e., $\hat{T}(1,0,t)$ and $\hat{T}(-1,0,t)$. (c) Comparison of the CSG thermal model against FEM and TEC models under the scenario of all-tabs and surface cooling (aTSC). (d) Maximum error of the CSG model with different orders under all cooling cases.}
    \label{Fig: FEM_MaxError}
\end{figure}

The examination of the developed modelling method and their underlying assumptions for control-oriented applications is conducted for the battery cell and cooling scenarios specified in Section~\ref{Sec: Sim setup}. Different CSG model orders, including $O$ = 1, 4, 9, 16, and 25, were evaluated. The results for cylindrical cell modelling compared with the ground truth represented by the FEM model are presented in Fig.~\ref{Fig: FEM_MaxError}. 

According to Fig.~\ref{Fig: FEM_MaxError}(d), the developed CSG model with an order of 25 can almost perfectly capture the battery cell's distributed thermal dynamics under any of the considered cooling scenarios. The resulting modelling error is bounded by 0.03~$^\circ$C. Reducing the model order from 25 to 9, the maximum temperature deviation between the model's output and its ground truth is 0.4~$^\circ$C within the entire test process. Similar observations are obtained for the studied pouch cell, in which the prediction error for the model with an order of 9 is less than 0.6~$^\circ$C. Remarkably, this high level of model fidelity is achieved in the presence of aggressive driving and cooling. 
The prediction error gradually increases when the model order is further lowered. This is consistent with {\it Remark \ref{rem:model-error}} and the applied finite-sum approximations in \eqref{Eq: T_h approx} and \eqref{Eq: Tp_hat}. Indeed, the model with an order of 1 or 4 can still well follow the trajectories of the true surface and core temperatures under different cooling scenarios, as seen in Fig.~\ref{Fig: FEM_MaxError}(b) and (c). Specifically, for SC, which involves one-side cooling, $O$ = 1 results in a maximum error of 1.26~$^\circ$C. For aTSC, which involves resolution in multiple directions (multi-side cooling), the same order produces a maximum error of 2.10~$^\circ$C.

 The impact of current rates on model accuracy is also evaluated. To do so, the mWLTP profile is scaled by factors of two and three, resulting in maximum C-rates of 2C and 3C, respectively. Under these highly aggressive operating conditions and the SC scenario, the errors of the CSG model with different orders are summarized in Table~\ref{tab: error}. 
The results demonstrate a clear necessity to increase the model order for accurately capturing battery cell thermal dynamics at elevated current rates due to pronounced thermal gradients and intensified spatial interactions.
Considering a maximum permissible error of 1.5~$^\circ$C, 
model orders $O$ = 1, 9, and 16 are sufficient to faithfully model the thermal behaviour in the mWLTP, mWLTP$\times$2, and mWLTP$\times$3 scenarios, respectively. It is worth mentioning that with this relatively small magnitude of local errors obtained under all considered cooling scenarios and extreme battery usage conditions, the CSG model proves suitable for a wide range of battery management applications.

\begin{table} [tbh]
  \caption{Effect of current rates on model accuracy.}
 \begin{center}
\begin{tabular}{c|c|c|c|c|c}
\hline
{Current profile}         & $O$=1 & $O$=4 & $O$=9 & $O$=16 & $O$=25      \\
\hline 
{mWLTP}           & 1.26 &  0.46  & 0.13  & 0.09   & 0.03 \\
{mWLTP$\times$2}  & 3.83  &  2.36  & 1.11  & 0.98   & 0.73  \\
{mWLTP$\times$3}  & 4.89  &  3.62   & 1.85   & 1.36    & 1.01 \\
\hline
    \end{tabular}
     \label{tab: error}
  \end{center}
\end{table}


The above results consequently verify the effectiveness of the modelling approach proposed in Section~\ref{Sec: Model reformulation}. Specifically, this confirms that the developed adaptable approach for the particular solution $T_p$ can efficiently handle any abrupt changes in the operating conditions and spatial gradients distributed across the battery cell; the CSG method can well reproduce the state dynamics governed by the PDE and its homogeneous solution. Furthermore, it numerically proves the conclusions drawn in {\it Remark~\ref{rem:superposition}}. Namely, for the particular solution problem in \eqref{Eq:ScaledBCs_Tp}, the response to the cooling power applied at individual boundaries satisfies the superposition principle and can be expressed as a linear combination of the cooling power applied at these boundaries.


In addition to the above validation against the high-fidelity FEM model implemented in COMSOL, our future work will explore direct experiments to evaluate our model under various cooling configurations. 

\subsection{Results of comparison against the TEC model}
\begin{figure}[ht]
    \centering
\includegraphics[width=1.0\linewidth]{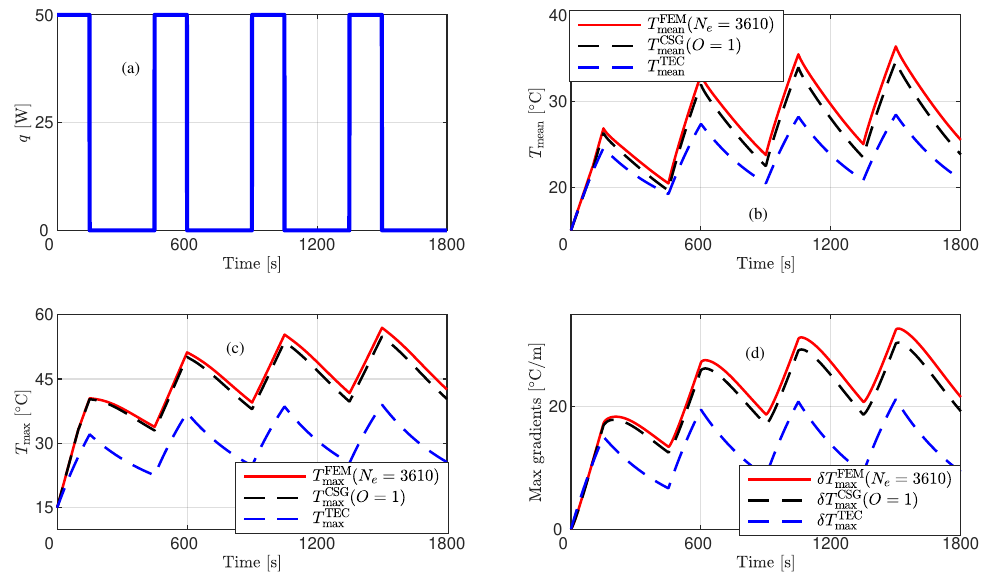}
    \caption{Comparison between the CSG and TEC models under surface cooling. (a) Pulse heat generation profile. (b) Evaluation of the mean temperature responses of the two models, i.e.,  $T^{\text{CSG}}_{\text{mean}}$ and  $T^{\text{TEC}}_{\text{mean}}$, against their ground truth. 
    (c) Comparison of their max temperature responses, which are $T^{\text{CSG}}_{\text{max}}$ and $T^{\text{TEC}}_{\text{max}}$, (d) Maximum thermal gradients of the CSG  and TEC models, i.e., $\delta T^{\text{CSG}}_{\text{max}}$ and $\delta T^{\text{TEC}}_{\text{max}}$.}
    \label{Fig: TEC_CSG}
\end{figure}

To investigate the efficacy and applicability of the CSG model validated in Section~\ref{Sec: Validation against FEM}, we compare it with the second type of benchmark defined in Section~\ref{sec:benchmark}, i.e., the TEC model. In the TEC model, the average temperature is derived as $T^{\text{TEC}}_{\text{mean}} = (T_{s}+T_{c})/2$, and the radial gradient is defined by $\delta T^r_{max}=(T_c - T_s)/(R_\text{out} - R_\text{in})$. To possibly fully reveal their predictive capability, particularly under highly dynamic operating conditions, we introduce a pulsed power load profile to generate the heat $q(t)$, as shown in Fig.~\ref{Fig: TEC_CSG}(a). This heat trajectory is characterised by high heat generation rates but a relatively short duration. For completeness, the TEC model was also included in the WLTP validation scenario shown in Fig.~\ref{Fig: FEM_MaxError}(b) and (c). The comparative results between the FEM, CSG, and TEC models under the pulsed power profile with surface cooling are presented in Fig.~\ref{Fig: TEC_CSG}. 

It is found that the CSG model of $O$ = 1 is capable of closely following the true trajectories of the mean temperature, max temperature, and max gradient in the entire test cycle. By contrast, in general, the TEC model with an order of 2 significantly underestimates all three considered indicators, resulting in temperature errors of several degrees. 
To observe more closely, the model's error for $T_{\max}$ is much larger than that for $T_\text{mean}$, and it increases when there are abrupt changes in the heat generation. This observation is consistent with the comparison under the WLTP profile in Fig.~\ref{Fig: FEM_MaxError}(b) and (c), where the TEC model similarly underestimates the core and surface temperatures. This means that compared to our developed model, the TEC model is unreliable for predicting the local thermal behaviour and may be limited to applications with only low heat generation in the battery cell. Particularly, this model cannot be used for battery safety management, where the maximum temperature must be accurately estimated and constrained, and for thermal uniformity control, where the information on the maximum gradients is critical. 

Computational efficiency is of great importance for model-based applications, particularly real-time thermal optimization and control of large-scale battery packs. With this in mind, we examine the computational times of the developed model with various orders and compare them with that of the TEC and FEM models. To have a fair comparison, we run each model ten times and compute their average simulation times in milliseconds (ms), with the results listed in Table~\ref{tab: compu-time}. Clearly, the CSG model with an order $O$=1 is more efficient than its TEC counterpart, leading to a reduction of 28.7\% in the required time. The model with higher orders takes longer to simulate, but its time consumption remains within the same order of magnitude as that of the TEC model. As expected, the FEM model takes a longer computational time, about 8s to compute the solution, making it impractical for real-time control applications.

The above results indicate that the CSG model with $O$=1 offers distinct advantages in both accuracy and computational efficiency. Therefore, it can readily replace the TEC model in existing BMSs to improve battery thermal performance in the real world. In addition, it may be computationally feasible to use a high-order CSG model, e.g., the order of 4, within a BMS, allowing a significantly more advanced management of battery safety and lifetime.   
\begin{table} [tbh]
  \caption{Average computational times of CSG model of different orders compared with the TEC model.}
 \begin{center}
\begin{tabular}{c|c|c|c|c|c|c|c}
\hline
Model type  & FEM  & TEC & $O$=1 & $O$=4 & $O$=9 & $O$=16 & $O$=25 \\
\hline 
Time [ms] & 8000 & 307 & 219 & 431 & 517 & 652  & 746  \\
\hline
    \end{tabular}
     \label{tab: compu-time}
  \end{center}
\end{table}

\section{Model-based applications} \label{Sec: model-based}
The developed models for cylindrical and pouch battery cells, featuring high fidelity and efficiency, can be effectively applied to various model-based applications. These include rapid analysis of different cooling scenarios, real-time closed-loop temperature control, and thermal optimization of cell design, as demonstrated in this section. Here, a model order $O=9$ is selected and simulations are conducted under the mWLTP profile on the battery cell and cooling scenarios specified in Section IV-B.

\subsection{Model-based analysis for different cooling scenarios} \label{Sec: analysis cooling cases}
For a given battery system and usage profiles, selecting the most appropriate cooling scenario can make a big difference to all its key properties. We analyze thermal performance merits, $T_{\text{mean}}$, $T_{\max}$, $\delta T^r_{\max}$, $\delta T^z_{\max}$, and $\Delta T = T_{\max}-T_{\min}$, under the five cooling scenarios defined in Section~\ref{Sec: Sim setup}, with the results for cylindrical cells presented in Table~\ref{tab: summary 5cases}. 

In terms of $T_{\text{mean}}$, bTC gives rise to the highest value at 22.77~$^\circ$C, followed by SC with the second-highest value. This is because the average temperature is primarily influenced by the cooling area, and both bTC and SC have only one side exposed to cooling. With the largest cooling area, aTSC demonstrates a high capability of heat removal, leading to the lowest values of $T_{\text{mean}}$ and $T_{\max}$ among all cases. However, aTSC does not achieve the lowest thermal gradient. In contrast, btTC provides the lowest $\delta T^r_{\max}$ at the local level and the lowest $\Delta T$ across the entire cell. This can be attributed to symmetric cooling from the cell's top and bottom sides, and that all electrodes get access to cooling at the same time under btTC because of the rolled electrode layer structure. On the contrary, in the case of SC, the outer layers get cooled before the internal layers, leading to inhomogeneous cooling and consequently the highest gradient of 32.1~$^\circ$C/mm. 

In all scenarios, it is observed that the cylindrical cell experiences significantly higher thermal gradients along the radial direction compared to the axial direction. This is due to the radial thermal conductivity being two orders of magnitude lower than that of the axial direction, creating a significant bottleneck in heat transfer to the surface. This, combined with the absence of cooling applied to the core, explains why the core temperature is the highest. Increasing the cooling flow rate and using battery materials with higher thermal conductivity can enhance cooling efficiency and help reduce the core temperature.
\begin{table} [tbh]
  \caption{Thermal performance merits under the five scenarios.}
  \vspace{-0.5cm}
 \begin{center}
\begin{tabular}{c|c|c|c|c|c}
\hline
  Scenarios & $T_{\text{mean}}$ & $T_{\max}$ & $\delta T^r_{\max}$ & $ \delta T^z_{\max}$  & $\Delta T$ \\
\hline
SC & 20.80 & 51.00 & 32.10 & {\bf 0.44} &36.00 \\
bTC & 22.77 & 54.01 & 14.05 & 8.94  & 39.01 \\
bTSC &  19.38 & 46.18 & 27.55 & 6.70 & 31.22  \\
btTC & 20.29 & 43.36 & {\bf 9.32} & 3.40 & {\bf 23.37} \\
aTSC & {\bf18.30} & {\bf 39.68} & 21.42 & 3.17 & 24.70 \\
\hline
    \end{tabular}
     \label{tab: summary 5cases}
  \end{center}
\end{table}

Moreover, it is evident that no single cooling scenario consistently outperforms the others across all performance metrics. Cooling only one of the cell's tabs, as represented by bTC and bTSC, is unfavourable due to its low removed heat and the significant $\delta T^z_{\max}$. When the cell's surface is involved in cooling, as in SC, bTSC, and aTSC, a considerable $\delta T^r_{\max}$ is observed. Among SC, btTC, and aTSC, each scenario performs the best in one or two specific metrics. For example, btTC offers the most homogeneous cooling effect but results in a slightly higher $T_{\text{mean}}$, whereas aTSC lowers $T_{\text{mean}}$ at the cost of increasing $\delta T^r_{\max}$. This indicates that different performance metrics may be in competition within each cooling scenario.

\subsection{Closed-loop temperature control} 
The performance trade-offs revealed in Section~\ref{Sec: analysis cooling cases} 
necessitate the selection of a scenario to satisfy application-specific requirements. In this regard, an interesting question is: which cooling scenario produces the lowest thermal gradient while maintaining a fixed average temperature? 

\begin{figure}[tbh]
    \centering
    \includegraphics[scale=0.44]{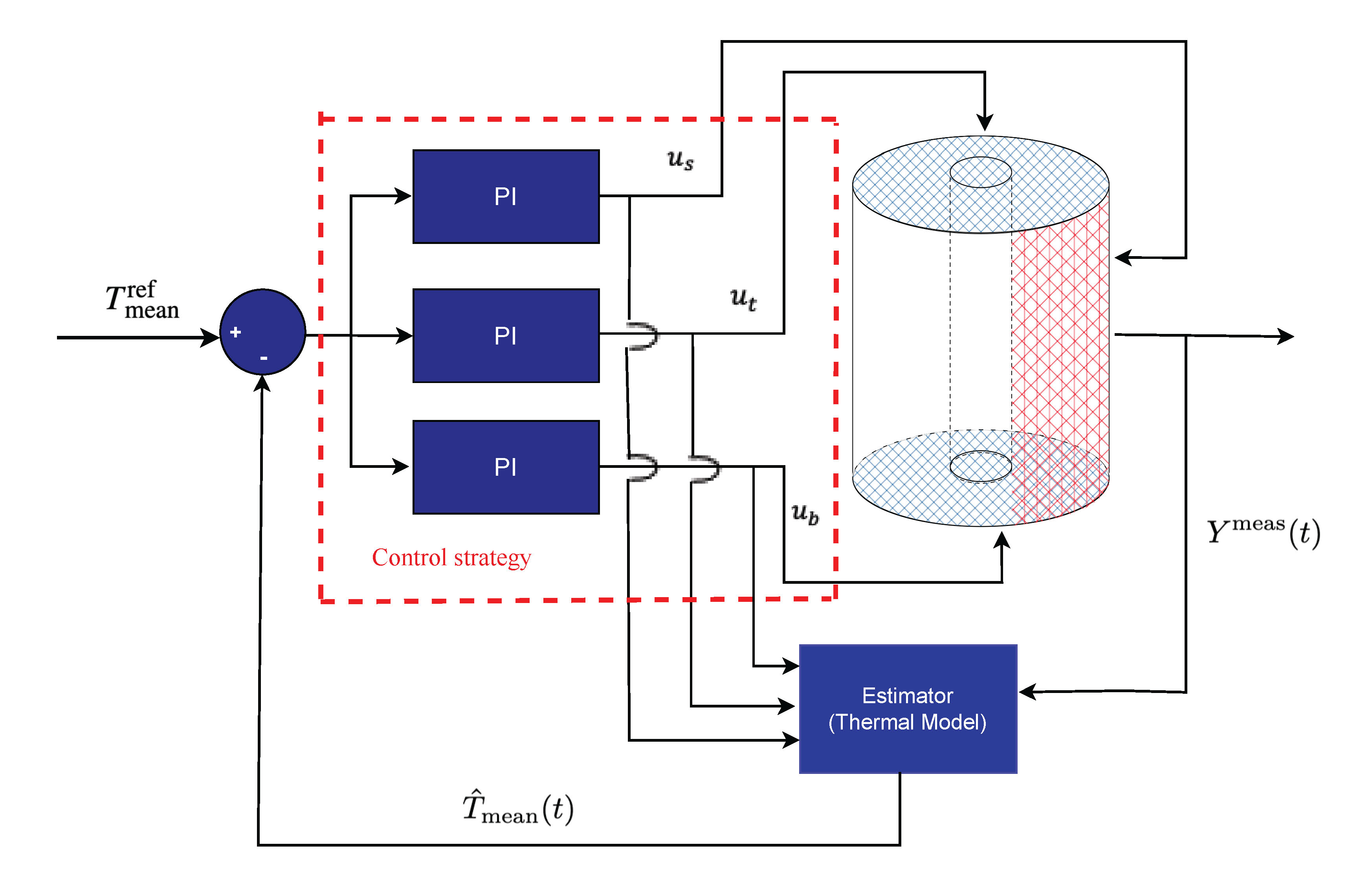}
    \caption{PI-based control scheme that tracks a given reference for the cylindrical cell's average temperature. $T^{\text{ref}}_{\text{mean}}$ and $\hat{T}_{\text{mean}}$ represent the set-point and estimate of the average temperature, respectively,
    and $Y^{\text{meas}}$ is the cell's measurable outputs. The cooling power applied to the surface, top, and bottom, which are $u_s$, $u_t$, and $u_b$ defined in \eqref{Eq: new controls}, is regulated by their corresponding PI controllers. Their design parameters are set based on the specific cooling scenario.}
    \label{Fig: PI}
\end{figure}

To answer this question and demonstrate the application of our developed battery thermal models, we implement a simple feedback control strategy based on multiple proportional-integral (PI) controllers, as framed in Fig.~\ref{Fig: PI}. This control scheme is designed to maintain the average cell temperature, $T_{\text{mean}}$, at a predefined set-point, $T^{\text{ref}}_{\text{mean}}$, which is specified as $20$~$^\circ$C. For each cooling scenario, the control in the framework of Fig.~\ref{Fig: PI} is tailored so that individual PI controllers regulate the cooling power directed to specific sides of the cell. For example, in the case of aTSC, all the top, bottom, and surface controllers are active, while for SC, only the surface controller is engaged, with $u_t(t)=u_b(t)=0$. To simplify the control implementation, we assume a uniform distribution of cooling power across all active areas.

Since $T_{\text{mean}}$ is not measurable during real-world battery operation, it is imperative to incorporate a state estimator within the designed control strategy. In this preliminary study, we design an open-loop estimator that mirrors the model \eqref{Eq:SS-model}--\eqref{Eq:SS-model-definition} for cylindrical battery cells. However, the state-space model allows any standard state-estimator such as the Kalman filter. Beyond its high accuracy and efficiency, this thermal model enables precise control of cooling applied to any side of the cell or combinations thereof, which is a capability not offered by existing models such as those in \cite{richardson2016board,hu2020control}. This flexibility makes our model particularly suitable for battery temperature estimation and control in sophisticated cooling scenarios.
\begin{figure}[ht]
    \centering    \includegraphics[width=1.0\linewidth]{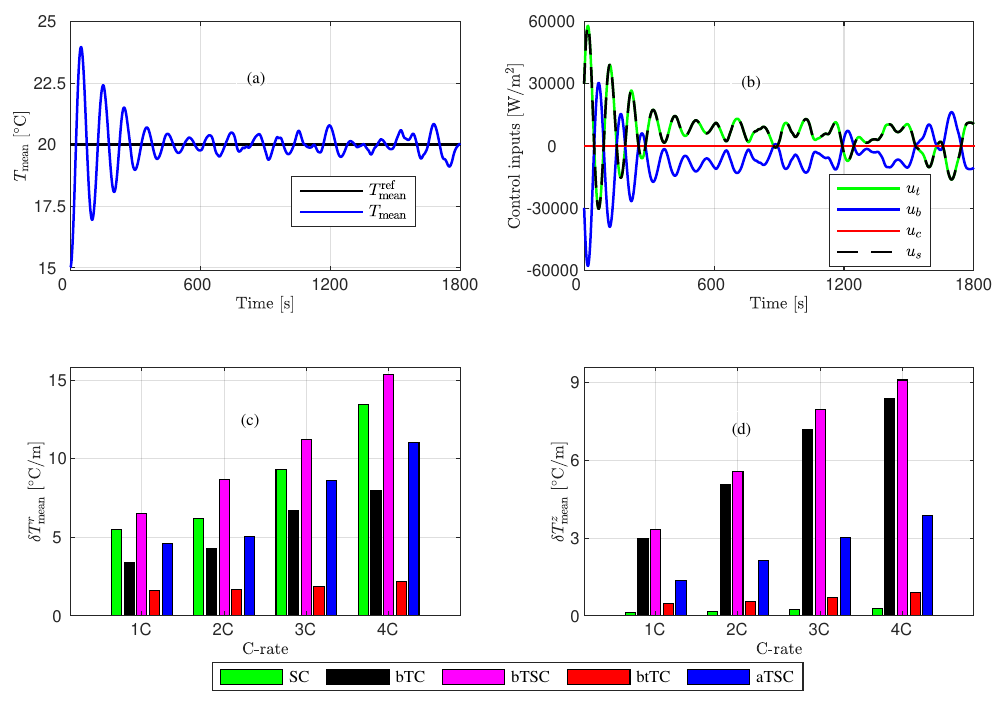}
    \caption{Results of battery thermal control of the cylindrical cell. (a) and (b) show the results obtained in aTSC and the 1C case of WLTP. 
    (c) and (d) compare the mean values of the radial and axial thermal gradients, respectively, for different cooling scenarios and WLTP cases.}
    \label{Fig: PI_results}
\end{figure}
For case studies, the original WLTP profile is scaled so that the maximum current becomes 1, 2, 3, and 4C, respectively. The control results are given in Fig.~\ref{Fig: PI_results}. To save space, in Fig.~\ref{Fig: PI_results}(a) and (b), only the trajectories obtained in aTSC and the 1C case are depicted.  

Based on the results in Fig.~\ref{Fig: PI_results}(c) and (d), one can readily find the answer to the question raised. When the average cell temperature is maintained at 20~$^\circ$C, btTC best balances the thermal gradients along the $r$- and $z$-directions and produces the most homogeneous thermal distribution across the cylindrical cell. Specifically, in btTC, the average thermal gradient along the $r$-direction, denoted by $\delta T^r_\text{mean}$, is substantially smaller than those observed in all other cooling scenarios. Regarding the average thermal gradient along the $z$-direction, btTC is comparable to the best-performing scenario, i.e., SC. 

In terms of the control's behaviour, as demonstrated in Fig.~\ref{Fig: PI_results}(a) and (b), substantial control efforts are initially required to attenuate the large tracking errors. Once $T_{\text{mean}}$ approaches the set-point, the magnitude of $u_t$, $u_b$, and $u_s$ decreases significantly. More sophisticated control strategies will be explored and implemented in future work to systematically enhance the control performance, in which multiple control objectives and potential constraints will be explicitly addressed. Within the control strategy, the new models developed in Section~\ref{Sec: Model reformulation} can serve as efficient tools for accurately simulating and predicting battery thermal states.

\subsection{Model-based evaluation and enhancement for cell design}
\begin{figure}[ht]
    \centering
    \includegraphics[width=1.0\linewidth]{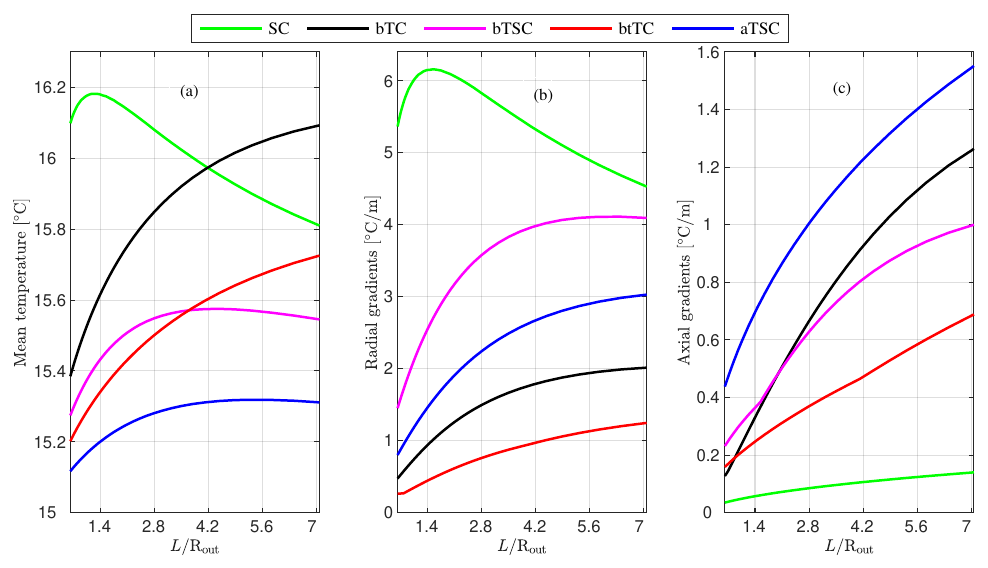}
    \caption{Evaluation of the thermal performance of cylindrical cells with varying dimensions under five cooling scenarios. (a) Evolution of the mean temperature $T_{\text{mean}}$. (b) Evolution of the maximal thermal gradients in the radial direction $\delta T^r_{\text{max}}$. (c) Evolution of the maximal thermal gradients in the axial direction $\delta T^z_{\text{max}}$.}
    \label{Fig: Cell_Dimensions}
\end{figure}

The dimensions of a battery cell are crucial in determining its thermal properties. Therefore, it is necessary to understand the impact of geometric parameters for optimized cell design. By maintaining a constant cell volume while varying the height-to-radius ratio, i.e., $L/R_\text{out}$, Fig.~\ref{Fig: Cell_Dimensions} illustrates the evolution profiles of the average temperature and maximum thermal gradients for cylindrical cells under the original WLTP. Note that with the cell volume being fixed, the heat generation trajectory for $q(t)$ remains unchanged regardless of varying dimensions. 

The results show that under the cooling scenarios bTC, bTSC, btTC, and aTSC, a higher value of $L/R_\text{out}$ typically leads to a more pronounced temperature rise and greater thermal gradients across the battery cell. This means that the cell's thermal behaviour is highly sensitive to its dimensions. Under these four scenarios, short and bulky battery cells are more favourable compared to their tall and slim alternatives. Specifically, the reduced mean temperature can help mitigate battery ageing mechanisms such as solid-electrode interface (SEI) growth, while the more uniform thermal distribution promotes more homogeneous usage of active electrode materials, both of which contribute to a longer cell lifespan \cite{lin2021review, zadeh2022thermal, xie2020mpc}. These findings can be used to evaluate existing cell designs. For battery types of 18650, 26650, 21700, and 4680 prevalent in today's market, their $L/R_\text{out}$ values are 7.22, 5.42, 6.67, and 3.48, respectively. It is evident that the 4680 cell type should offer the best thermal performance across the four scenarios involving at least one tab in the cooling process.

However, in the case of SC, the effect of cell dimensions becomes more complex. As $L/R_\text{out}$ grows, $T_{\text{mean}}$ and $\delta T^r_{\text{max}}$ show a slight increase when $L/R_\text{out}<1.4$ but rapidly decrease beyond the threshold of 1.4. Therefore, cell designs should be optimised based on the targeting cooling scenarios.  

\section{Conclusion} \label{Sec: conclusions}
This article has developed a new battery modelling framework for spatially-resolved thermal dynamics based on Chebyshev spectral-Galerkin approach and model component decomposition. The obtained library of models is well-suited for online thermal performance optimisation, thanks to its ability to independently control the tab and surface cooling channels of the battery. This framework has been presented for both cylindrical and pouch cells and thoroughly evaluated through various combinations of tab and surface cooling cases under real-world vehicle driving profiles. 

By using four states, the model accurately predicts the local thermal behaviour, even under aggressive vehicle driving and cooling scenarios. The model with only one state, in particular, is both more accurate and computationally efficient than the widely utilised two-state thermal equivalent circuit (TEC) model, with a 28.7\% reduction in computational time. Even as the number of states increases, it remains in the same computational order of magnitude as the TEC model, suggesting it could readily replace the TEC model in existing battery management systems. 

Model-based analysis on the cylindrical cell demonstrated that exclusively cooling through the tabs on the top and the bottom sides produced the lowest thermal gradients under various 
C-rates. Furthermore, the model's utility extends to the design optimisation of battery cells. Dimension perturbation analysis showed that a short and bulky cylindrical cell enhances thermal performance compared to its tall and slim counterpart, indicating that current market cell designs may not be ideal for thermal management. Beyond batteries, the versatile modelling framework can be extended to other systems governed by partial differential equations, making it a valuable tool across multiple engineering disciplines. 

\bibliographystyle{IEEEtran}%
\bibliography{references}%

\end{document}